\documentclass[%
 reprint,
superscriptaddress,
 amsmath,amssymb,
 aps,
floatfix
]{revtex4-2}

\usepackage{siunitx}

\usepackage{graphicx}
\usepackage{dcolumn}
\usepackage{bm}
\usepackage[pdfencoding=auto, psdextra, colorlinks=true]{hyperref}
\hypersetup{urlcolor=blue, citecolor=blue, linkcolor=blue, breaklinks=true}
\usepackage{rotating}
\usepackage{epstopdf}

\usepackage{silence}
\begin{document}

\title{Spectroscopic study of \texorpdfstring{$^{47}$Ca}{47Ca} from the \texorpdfstring{$\beta^-$}{beta-minus} decay of \texorpdfstring{$^{47}$K}{47K}}

\author{J.K.~Smith}
 \email{JSmith@pierce.ctc.edu}
 \altaffiliation[Present address: ]{Pierce College Puyallup, 1601 39th Ave SE, Puyallup, WA, 98374, USA}
 \affiliation{TRIUMF, 4004 Wesbrook Mall, Vancouver, BC, V6T 2A3, Canada}
 
\author{A.B.~Garnsworthy} 
\affiliation{TRIUMF, 4004 Wesbrook Mall, Vancouver, BC, V6T 2A3, Canada}

\author{J.L.~Pore} 
\altaffiliation[Present address: ]{Lawrence Berkeley National Laboratory, Berkeley, CA 94720, USA}
\affiliation{Department of Chemistry, Simon Fraser University, Burnaby, BC, V5A 1S6, Canada}

\author{C.~Andreoiu} 
\affiliation{Department of Chemistry, Simon Fraser University, Burnaby, BC, V5A 1S6, Canada}

\author{A.D.~MacLean} 
\affiliation{Department of Physics, University of Guelph, Guelph, ON, N1G 2W1, Canada}

\author{A.~Chester} 
\altaffiliation[Present address: ]{National Superconducting Cyclotron Laboratory, Michigan State University, East Lansing, MI 48824, USA}
\affiliation{TRIUMF, 4004 Wesbrook Mall, Vancouver, BC, V6T 2A3, Canada}

\author{Z. Beadle}
\affiliation{Reed College, 3203 Southeast Woodstock Boulevard, Portland, OR, 97202, USA}

\author{G.C.~Ball} 
\affiliation{TRIUMF, 4004 Wesbrook Mall, Vancouver, BC, V6T 2A3, Canada}

\author{P.C.~Bender} 
\altaffiliation[Present address: ]{Department of Physics and Applied Physics, University of Massachusetts Lowell, Lowell, MA, 01854, USA}
\affiliation{TRIUMF, 4004 Wesbrook Mall, Vancouver, BC, V6T 2A3, Canada}

\author{V.~Bildstein} 
\affiliation{Department of Physics, University of Guelph, Guelph, ON, N1G 2W1, Canada}

\author{R.~Braid} 
\affiliation{Department of Physics, Colorado School of Mines, Golden, CO, 80401, USA}

\author{A.~Diaz Varela} 
\affiliation{Department of Physics, University of Guelph, Guelph, ON, N1G 2W1, Canada}

\author{R.~Dunlop} 
\affiliation{Department of Physics, University of Guelph, Guelph, ON, N1G 2W1, Canada}

\author{L.J.~Evitts}
\altaffiliation[Present address: ]{Nuclear Futures Institute, Bangor University, Bangor, Gwynedd, LL57 2DG, United Kingdom}
\affiliation{TRIUMF, 4004 Wesbrook Mall, Vancouver, BC, V6T 2A3, Canada}
\affiliation{Department of Physics, University of Surrey, Guildford, Surrey, GU2 7XH, United Kingdom}

\author{P.E.~Garrett} 
\affiliation{Department of Physics, University of Guelph, Guelph, ON, N1G 2W1, Canada}

\author{G.~Hackman}
\affiliation{TRIUMF, 4004 Wesbrook Mall, Vancouver, BC, V6T 2A3, Canada}

\author{S.V.~Ilyushkin} 
\affiliation{Department of Physics, Colorado School of Mines, Golden, CO, 80401, USA}

\author{B.~Jigmeddorj} 
\altaffiliation[Present address: ]{Canadian Nuclear Laboratories, 286 Plant Rd, Chalk River, ON, K0J 1J0, Canada}
\affiliation{Department of Physics, University of Guelph, Guelph, ON, N1G 2W1, Canada}

\author{K.~Kuhn} 
\affiliation{Department of Physics, Colorado School of Mines, Golden, CO, 80401, USA}

\author{A.T.~Laffoley} 
\affiliation{Department of Physics, University of Guelph, Guelph, ON, N1G 2W1, Canada}

\author{K.G.~Leach} 
\altaffiliation[Present address: ]{Department of Physics, Colorado School of Mines, Golden, CO, 80401, USA}
\affiliation{TRIUMF, 4004 Wesbrook Mall, Vancouver, BC, V6T 2A3, Canada}

\author{D.~Miller} 
\affiliation{TRIUMF, 4004 Wesbrook Mall, Vancouver, BC, V6T 2A3, Canada}

\author{W.J.~Mills} 
\affiliation{TRIUMF, 4004 Wesbrook Mall, Vancouver, BC, V6T 2A3, Canada}

\author{W.~Moore} 
\affiliation{Department of Physics, Colorado School of Mines, Golden, CO, 80401, USA}

\author{M.~Moukaddam}
\affiliation{TRIUMF, 4004 Wesbrook Mall, Vancouver, BC, V6T 2A3, Canada}

\author{B.~Olaizola} 
\affiliation{Department of Physics, University of Guelph, Guelph, ON, N1G 2W1, Canada}

\author{E.E.~Peters} 
\affiliation{Department of Chemistry, University of Kentucky, Lexington, KY, 40506-0055, USA}

\author{A.J.~Radich} 
\affiliation{Department of Physics, University of Guelph, Guelph, ON, N1G 2W1, Canada}

\author{E.T.~Rand} 
\affiliation{Department of Physics, University of Guelph, Guelph, ON, N1G 2W1, Canada}

\author{F.~Sarazin} 
\affiliation{Department of Physics, Colorado School of Mines, Golden, CO, 80401, USA}

\author{C.E.~Svensson} 
\affiliation{Department of Physics, University of Guelph, Guelph, ON, N1G 2W1, Canada}

\author{S.J.~Williams} 
\affiliation{National Superconducting Cyclotron Laboratory, Michigan State University, East Lansing, MI, 48824, USA}

\author{S.W.~Yates} 
\affiliation{Department of Chemistry, University of Kentucky, Lexington, KY, 40506-0055, USA}
\affiliation{Department of Physics \& Astronomy, University of Kentucky, Lexington, KY, 40506-0055, USA}

\date{\today}

\begin{abstract}
The $\beta^-$ decay of $^{47}$K to $^{47}$Ca is an appropriate mechanism for benchmarking interactions spanning the $sd$ and $pf$ shells, but current knowledge of the $\beta^-$-decay scheme is limited. We have performed a high-resolution, high-efficiency study of the $\beta^-$-decay of $^{47}$K with the GRIFFIN spectrometer at TRIUMF-ISAC. The study revealed 48 new transitions, a more precise value for the $^{47}$K half-life (17.38(3)~s), and new spin and parity assignments for eight excited states. Levels placed for the first time here raise the highest state observed in $\beta^-$ decay to within 568(3) keV of the $Q$-value and confirm the previously measured large $\beta^-$-decay branching ratios to the low-lying states. Previously unobserved $\beta^-$-feeding to 3/2$^+$ states between 4.5 and 6.1~MeV excitation energy was identified with a total $\beta^-$-feeding intensity of 1.29(2)\%.  The sum of the $B(GT)$ values for these states indicates that the $1s_{1/2}$ proton hole strength near this excitation energy is comparable to the previously known $1s_{1/2}$ proton and neutron hole strengths near 2.6 MeV.
\end{abstract}

\pacs{Valid PACS appear here}
\maketitle



\section{\label{sec:intro}Introduction}
Recently, \emph{ab initio} techniques for the nuclear many-body problem have reached $Z=20$ and beyond
\cite{Hagen2012,Holt2012,Hergert2013,Soma2014,Holt2014,Jansen2016,Hagen2016,Stroberg2017,Morris2018,Gysbers2019}. 
During the development and testing of these methods, it is important to benchmark these models against nuclei close to stability. Such benchmarks provide important feedback on the necessary and most relevant physical processes. Reproducing the doubly magic nature of $^{48}$Ca, for example, required the inclusion of three-body forces \cite{Holt2012}. Accurate and precise nuclear structure data, such as level energies, spins and parities, as well as branching ratios are necessary for these benchmarking tests.


The achievable precision for many experimental measurements (e.g., excited state energy levels and half-lives) can easily outstrip the accepted precision of theoretical calculations. In these cases, increased statistics and improved sensitivities can be helpful in improving the accuracy of experimental measurements. A higher level of statistics opens up new avenues of analysis, such as angular correlation measurements, which can be used to assign, constrain, or correct spin and parity assignments. Increases in sensitivity can reduce possible systematic errors due to the accumulation of unobserved transitions.

Here, we report on the most sensitive study of the $\beta^-$ decay of $^{47}$K performed to date, using the excellent sensitivity of the GRIFFIN spectrometer and the high beam intensity provided by TRIUMF-ISAC \cite{Garnsworthy2019}.
As of the last evaluation of the available data, over 100 states have been identified in $^{47}$Ca \cite{Burrows2007}. Only 20 of these states, however, have firm spin and parity assignments. Previous studies of the $\beta^-$ decay of $^{47}$K identified the principal $\beta^-$-decay branches with detection limits down to $\gamma$-ray intensities of 1\% of the strength of the strongest transition, leaving a gap of just over 4\,MeV between the state with the highest observed energy and the $Q$ value \cite{Warburton1970,Alburger1984}.
There are seventeen excited states in $^{47}$Ca that were unobserved in previous $\beta^-$-decay studies \cite{Burrows2007} for which tentative spin and parity assignments would allow their population via allowed transitions from the ground state of $^{47}$K. Potentially unobserved, low-intensity transitions from these states could change the calculated $\beta^-$ branching ratios significantly.
This lack of firm data has up to now limited the usefulness of this nuclide for tests of nuclear structure models.


\section{\label{sec:background}Previous measurements}
The results of previous measurements related to the $\beta^-$ decay of $^{47}$K are summarized here for context and referenced as needed during the later discussion of the current results.

The ground state of $^{47}$Ca has been clearly identified as 7/2$^-$ \cite{Hanspal1985}, and the first excited state as 3/2$^-$ \cite{Martin1972,Williams-Norton1977}. Just below 2600~keV lie two states that have been assigned spin-parities of 1/2$^+$ and 3/2$^+$. The location of $^{47}$Ca as one neutron removed from the closed $N=28$ means that these states are commonly identified as single-particle or single-hole states in the $0f_{7/2}$, $1p_{3/2}$, $1s_{1/2}$, and $0d_{3/2}$ orbitals, respectively. Above these energies, there are few states with firm spin-parity assignments, but more than 80 states that are tentative or constrained by transfer reaction data.

Three studies of the $\beta^-$ decay of $^{47}$K have previously been reported \cite{Kuroyanagi1964,Warburton1970,Alburger1984}.
In the first study, a source of $^{47}$K was created by the bombardment of $^{48}$Ca with bremsstrahlung \cite{Kuroyanagi1964}. $\gamma$ rays at energies of 2.0 and 2.6~MeV were observed, and the half-life of $^{47}$K was measured as $t_{1/2}=17.5(3)$~s \cite{Kuroyanagi1964}. Warburton {\it et al.} \cite{Warburton1970} produced $^{47}$K by the bombardment of $^{48}$Ca with tritons and identified $\gamma$ rays with energies of 2013.13(30), 564.74(30), and 585.75(30)~keV. They did not observe a 2.6~MeV $\gamma$ ray, but attributed the $\gamma$ ray seen at this energy by Kuroyanagi {\it et al.} \cite{Kuroyanagi1964} as a sum of the 2013.13~keV $\gamma$ ray with the 564.74 or 586.75~keV $\gamma$ rays. A second study with improved energy resolution and $\gamma$-ray detection efficiency by the same group was published later and continues as the most precise measurement of $^{47}$K $\beta^-$ decay to date, with relative $\gamma$ intensities down to 1.12\% (if the 2013.13~keV $\gamma$ ray has an intensity of 100\%) \cite{Alburger1984}. At this time, three $\beta^-$ branches were identified: 81.0(15)\% to the 1/2$^+$ state, 19.0(3)\% to the 3/2$^+$ state, and $<2.0$\% to the 3/2$^-$ state, all populated from the 1/2$^+$ ground state of $^{47}$K.

In addition to $\beta^-$-decay studies, the known energy levels have been expanded greatly by experiments populating $^{47}$Ca from the nearby stable $^{48}$Ca \cite{Martin1972,Williams-Norton1977,Fortier1978,Kovar1978,Humanic1982,Petersen1983,Mando1983,Hanspal1985,Strauch2000,Broda2001,Schweda2001,Montanari2012,Crawford2017} and $^{46}$Ca \cite{Bjerregaard1965,Belote1966,Cranston1970}. The most recent NNDC evaluation lists 66 excited states below the $\beta^-$-decay $Q$-value of 6632.4(26)~keV \cite{Wang2017}; six of these have firm spin-parity assignments \cite{Burrows2007}. Most of these measurements were made without $\gamma$-ray detection, and the number of $\gamma$ rays assigned to $^{47}$Ca is limited to 19 \cite{Burrows2007,Crawford2017}. Three lifetime measurements have been made: 6.1(3) ps, 9.6(14) ps, and 6.8(6) ps for the 3/2$^-$, 3/2$^+$, and 1/2$^+$ states, respectively \cite{Montanari2012}.

\section{\label{sec:expsetup}Experimental setup}
The isotope $^{47}$K was produced following the bombardment of a multi-layered UC$_x$ target by a 500~MeV proton beam delivered by the TRIUMF Cyclotron \cite{Bylinskii2013} at a beam current of 10~$\mu$A. The surface-ionized $^{47}$K ions were extracted and accelerated to 28~keV before being selected by a high-resolution mass separator for delivery to the experiment station at a beam rate of $\sim$50,000 ions/s.

The ions were stopped in a Mylar tape, where the nuclei subsequently decayed. The radiation emitted was detected by the surrounding three detector systems: the Gamma-Ray Infrastructure For Fundamental Investigations of Nuclei (GRIFFIN), an array of high-purity germanium (HPGe) clovers for detection of $\gamma$ rays  \cite{Svensson2013,Rizwan2016,Garnsworthy2017,Garnsworthy2019}; SCEPTAR, an array of plastic scintillator paddles for the detection of $\beta$ particles; and PACES, an array of five lithium-drifted silicon crystals for the detection of conversion electrons. To accommodate the inclusion of PACES, only half of the SCEPTAR array was installed, and one of the 16 GRIFFIN clovers was removed. Each GRIFFIN clover was positioned at a source-to-detector distance of 11~cm from the implantation point. A sphere of 20~mm thick Delrin plastic absorber was placed around the vacuum chamber to stop energetic $\beta$ particles from reaching the HPGe detectors.

A tape system was used to collect data in a series of cycles. Each cycle included the following sections: background measurement (10~s), source accumulation (600 or 120~s), source decay (6 or 350~s), and source removal (1~s). This cycling allowed the periodic removal of the long-lived $^{47}$Ca daughter ($t_{1/2}$ = 4.536(3)~days, \cite{Burrows2007}). The strongest contaminant was the $\beta^-$ decay of $^{47}$Ca, whose activity was approximately $0.06\%$ the activity of $^{47}$K. Data were collected for approximately 27 hours.

Energy and timing signals were collected from each detector using an early implementation of the GRIFFIN digital data acquisition system \cite{Garnsworthy2017}, operated in a triggerless mode. The timing for each signal was determined with a digital leading-edge discriminator algorithm and tagged with a timestamp in 10~ns increments.

\section{\label{sec:results}Experimental methods and results}

\subsection{\label{ssec:gammaspec}\texorpdfstring{$\gamma$}{gamma}-ray spectra}

\begin{figure*}
	\centering
    \includegraphics[width=\textwidth]{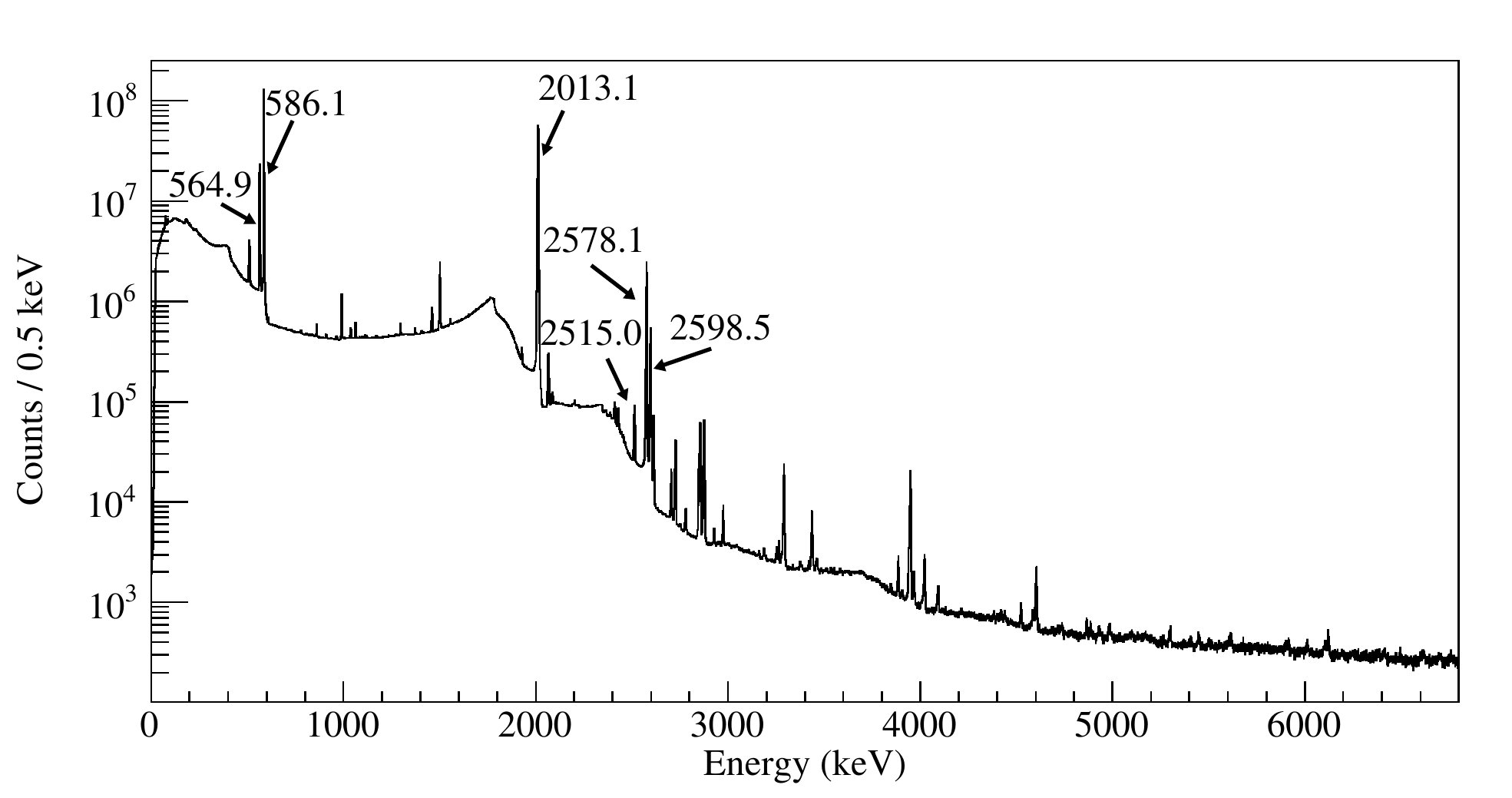}
	\caption{The $\gamma$-ray energy spectrum (using an addback algorithm) for the $\beta^-$ decay of $^{47}$K shows many $\gamma$ rays previously unseen in $\beta^-$-decay studies. The $Q$ value for this decay is 6632.4(26)~keV. The six $\gamma$ rays seen in previous $\beta^-$-decay studies are labeled.}
	\label{fig:singles}
\end{figure*}

\begin{figure*}
	\centering
    \includegraphics[width=\textwidth]{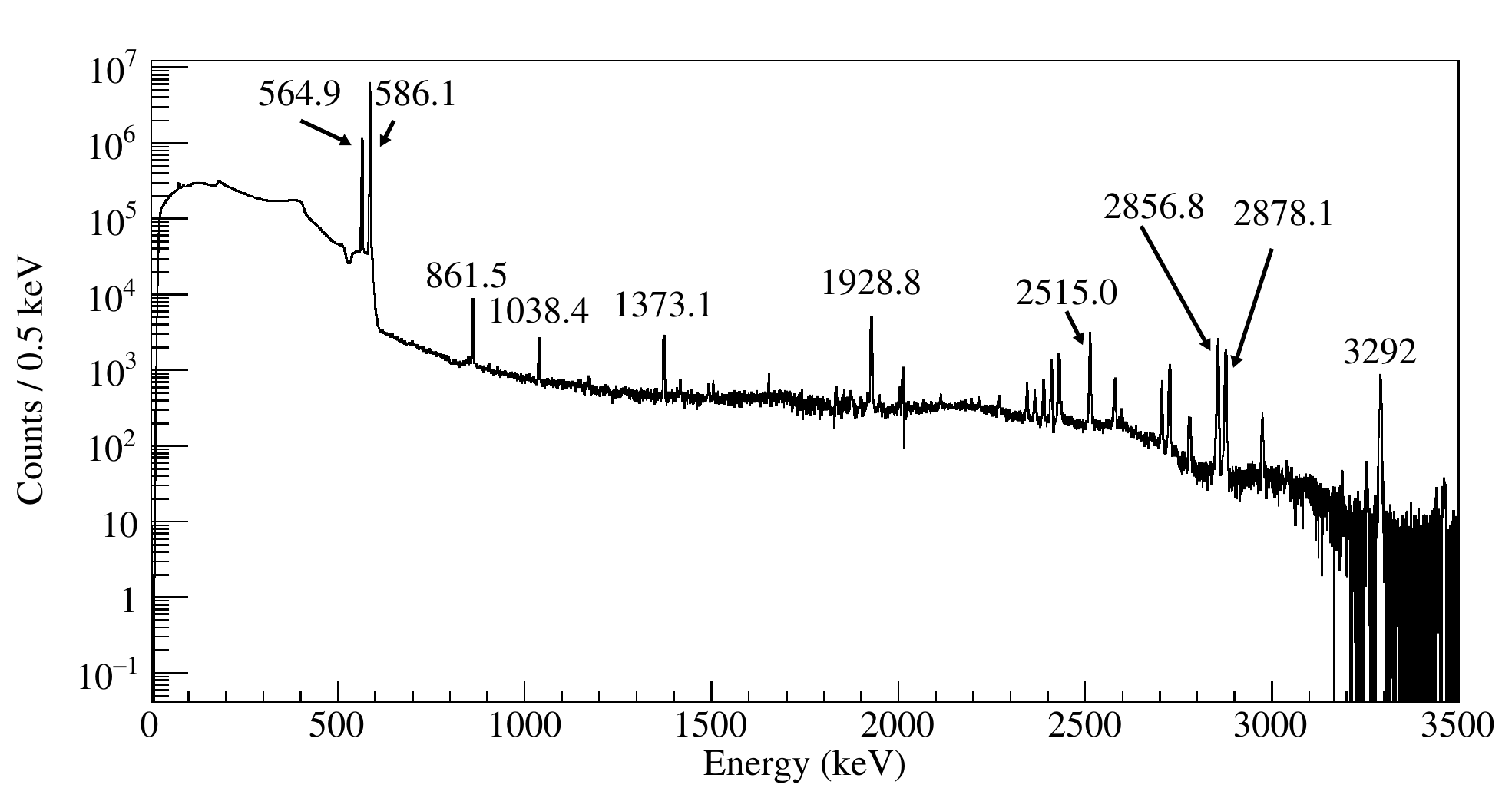}
    \caption{Energies of $\gamma$ rays observed in coincidence with a 2013.1~keV $\gamma$ ray.}
   	\label{fig:slices}
\end{figure*}

The energy signal of each HPGe crystal was calibrated with sources of $^{60}$Co, $^{152}$Eu, $^{133}$Ba, and $^{56}$Co. These sources provide a calibration up to 3500~keV with a precision of 0.3~keV. Higher energy $\gamma$ rays that should appear equal to the sum of two lower-energy $\gamma$ rays from a competing decay pathway were observed to have a measured energy systematically lower than the sum value. To account for this effect, the errors for energies above 3500~keV were inflated to 1~keV. Of the fifteen clover detectors present in the experiment, four crystals, in four different clovers, had poor energy resolution and were excluded from the rest of the analysis.

The bulk of this analysis was performed with signals from single crystals, but when measuring low-intensity $\gamma$ rays (and especially those at higher energies), an addback algorithm was employed to reduce the effect of the Compton continuum.
This algorithm combined all HPGe interactions within the same clover that occurred within 200~ns.
Fig. \ref{fig:singles} shows the $\gamma$-ray energy spectrum for the decay of $^{47}$K to levels in $^{47}$Ca with this addback algorithm applied.

For energies below 1500~keV, the efficiency of the full array was calibrated with $^{60}$Co, $^{152}$Eu, and $^{133}$Ba sources, with the appropriate corrections for summing. Above 1500~keV, the shape of the efficiency curve was determined by a $^{56}$Co source and GEANT4 simulations of the array that were scaled to match the measured efficiency at lower energies. The absolute efficiency at 1332~keV was found to be 7.2(1)\% without the addback algorithm and 10.1(1)\% with the addback algorithm.

Since the data acquisition system was operated in triggerless mode, coincident events were constructed offline as part of the analysis.
A matrix of coincident (defined as $\gamma$-rays with time differences less than 0.2 $\mu$s) $\gamma$-ray energies was constructed.
Included in this $\gamma$-$\gamma$ matrix of coincident $\gamma$-ray interactions is a small amount of time-random background. To remove this, a similar $\gamma$-$\gamma$ matrix was constructed from background events (with $\gamma$-$\gamma$ time differences between 0.6 and 1.4 $\mu$s), scaled, and subtracted from the prompt $\gamma$-$\gamma$ matrix to create a random-time corrected $\gamma$-$\gamma$ matrix. Each individual slice also had an energy-dependent background subtraction applied, with the normalization based on the spectral shape at nearby energies. Figure \ref{fig:slices} shows a slice from that matrix: $\gamma$ rays in coincidence with a 2013.1~keV $\gamma$ ray. For measurements of low-intensity $\gamma$ rays, a second matrix was constructed with the same process, but using $\gamma$-ray energies from the addback algorithm. 

\begin{sidewaysfigure*}[p]
	\centering
	\vskip0.33\linewidth
    \includegraphics[width=0.98\linewidth]{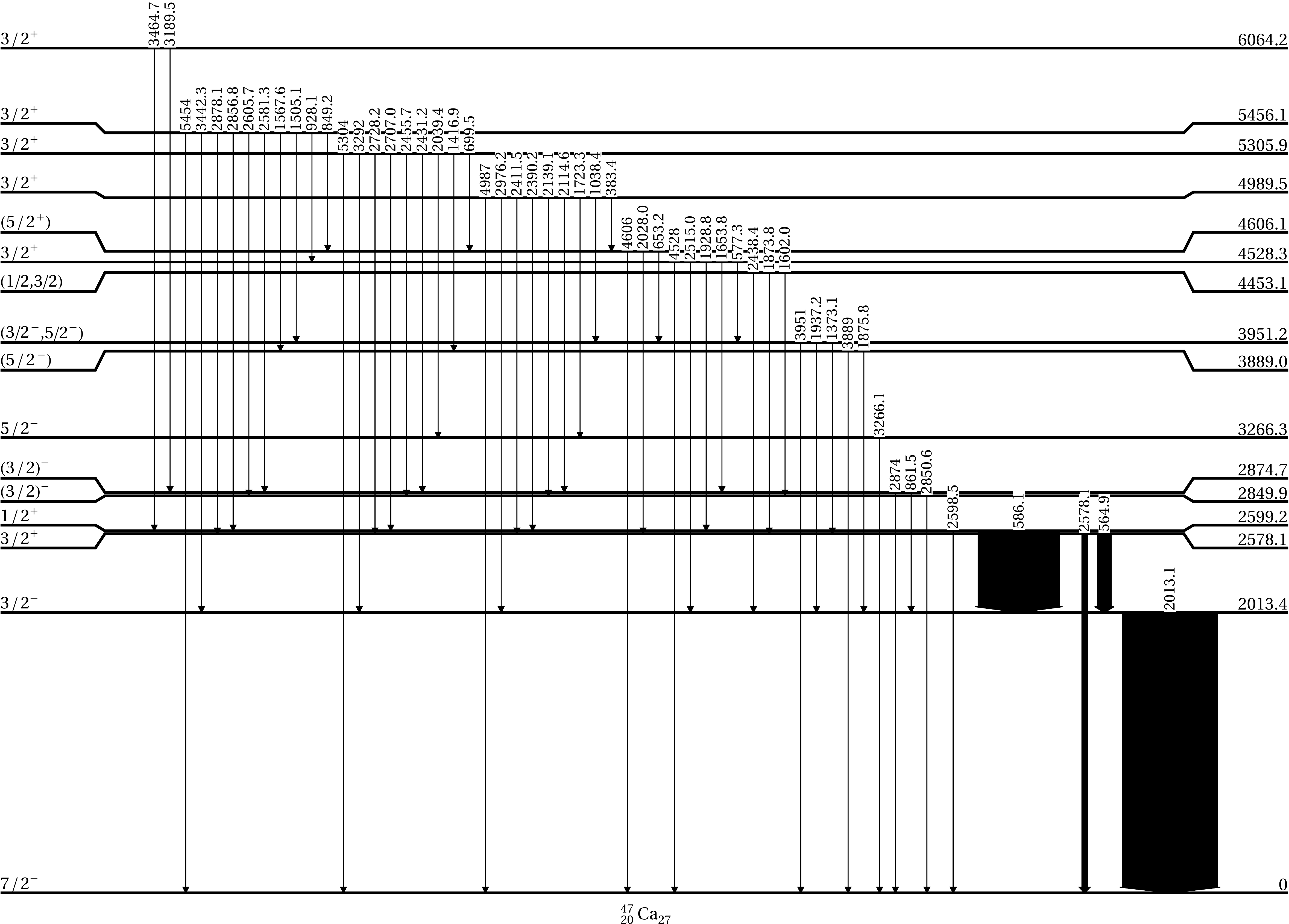}
    \caption{\label{fig:levelscheme1}Levels of $^{47}$Ca observed in the $\beta^-$ decay of $^{47}$K. The thickness of each arrow represents the relative intensity of that $\gamma$ ray. All transitions from states above 2600~keV are newly assigned within the $\beta^-$ decay level scheme.}
\end{sidewaysfigure*}

\begin{sidewaysfigure*}
	\centering
    \vskip0.33\linewidth
    \includegraphics[width=0.98\linewidth]{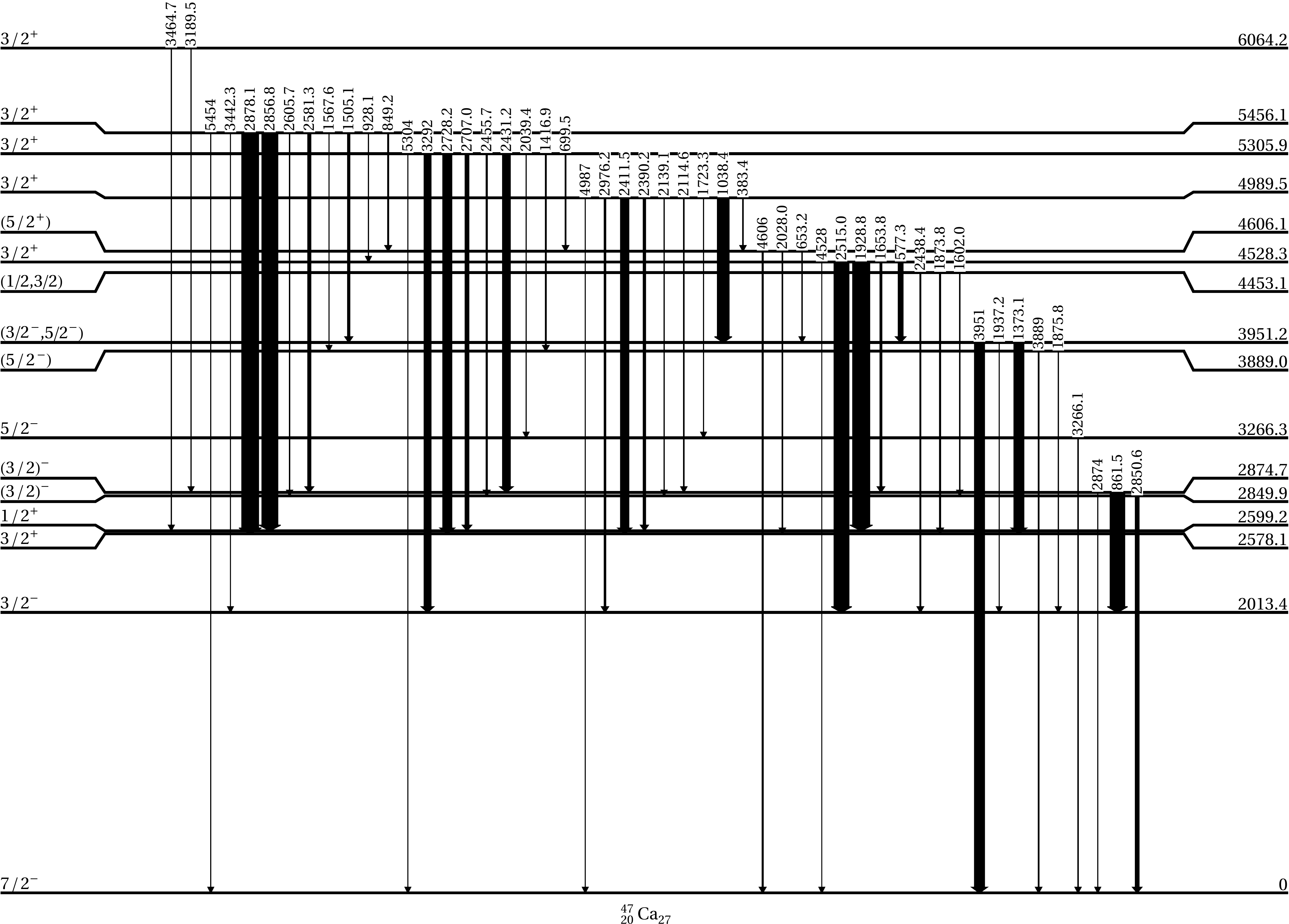}
    \caption{Levels of $^{47}$Ca observed in the $\beta^-$ decay of $^{47}$K. Only transitions from states above 2600~keV are shown. The thickness of each arrow represents the relative intensity of that $\gamma$ ray. All transitions shown are newly assigned within the $\beta^-$ decay level scheme.}
    \label{fig:levelscheme2}
\end{sidewaysfigure*}

Figs. \ref{fig:levelscheme1} and \ref{fig:levelscheme2} and Table \ref{table:intensities} show the $^{47}$Ca level scheme constructed with this $\gamma$-$\gamma$ matrix, with the spins and parities assigned in this work. The widths of the decay arrows are proportional to the intensities of the $\gamma$ rays relative to the 2013.1~keV $\gamma$-ray intensity. Fig. \ref{fig:levelscheme1} shows the full decay scheme, which is dominated by decays from states below 2600~keV whose relative intensities range from 0.38(3)\% to 100\%. In contrast, Fig. \ref{fig:levelscheme2} only shows decays from states above 2600~keV, which range in relative intensity from 0.00024(8)\% to 0.179(5)\%.

\subsection{\label{ssec:gintensities}\texorpdfstring{$\gamma$}{gamma}-ray relative intensities}

\begin{table*}
    \caption{\label{table:intensities}Data for states and $\gamma$ rays measured in the $\beta^-$ decay of $^{47}$K. Columns are: initial state energy ($E_i$), initial state spin ($J_i^{\pi}$), final state energy ($E_f$), final state spin ($J_{f}^{\pi}$), $\gamma$-ray energy ($E_{\gamma}$), $\gamma$-ray branching ratio ($BR_{\gamma}$), and $\gamma$-ray relative intensity ($I_{\gamma}$).}
	\begin{ruledtabular}
    \begin{tabular}{ccccccc}
	$E_i$ (keV)&$J_i^{\pi}$&$E_f$ (keV)&$J_f^{\pi}$&$E_\gamma$ (keV)&$BR_{\gamma}$&$I_{\gamma}$\\
\hline
2013.4(3)&3/2$^-$&0.0&7/2$^-$&2013.1(4)&100&100\\
2578.1(3)&3/2$^+$&0.0&7/2$^-$&2578.1(3)&39.9(7)&5.8(2)\\
&&2013.4(3)&3/2$^-$&564.9(3)&100.0(7)&14.5(2)\\
2599.2(3)&1/2$^+$&0.0&7/2$^-$&2598.5(4)&0.441(7)&0.381(8)\\
&&2013.4(3)&3/2$^-$&586.1(3)&100(2)&86(2)\\
2849.9(3)&(3/2)$^-$&0.0&7/2$^-$&2850.6(3)&100&0.0392(8)\\
2874.7(3)&(3/2)$^-$&0.0&7/2$^-$&2874(1)&0.88(8)&0.0013(1)\\
&&2013.4(3)&3/2$^-$&861.5(3)&100.00(8)&0.153(3)\\
3266.3(3)&5/2$^-$&0.0&7/2$^-$&3266.1(3)&100&0.0053(2)\\
3889.0(3)&(5/2$^-$)&0.0&7/2$^-$&3889(1)&100(3)&0.0076(4)\\
&&2013.4(3)&3/2$^-$&1875.8(3)&39(3)&0.0030(3)\\
3951.2(3)&(3/2$^-$,5/2$^-$)&0.0&7/2$^-$&3951(1)&100(1)&0.105(3)\\
&&2013.4(3)&3/2$^-$&1937.2(4)&0.56(8)&0.00059(9)\\
&&2578.1(3)&3/2$^+$&1373.1(3)&99(1)&0.104(2)\\
4453.1(4)&(1/2,3/2)&2013.4(3)&3/2$^-$&2438.4(4)&67(6)&0.013(1)\\
&&2578.1(3)&3/2$^+$&1873.8(4)&100(6)&0.019(3)\\
&&2849.9(3)&(3/2)$^-$&1602.0(3)&35(2)&0.0066(9)\\
4528.3(3)&3/2$^+$&0.0&7/2$^-$&4528(3)&0.13(10)&0.00023(18)\\
&&2013.4(3)&3/2$^-$&2515.0(3)&89(2)&0.158(6)\\
&&2599.2(3)&1/2$^+$&1928.8(3)&100(2)&0.179(4)\\
&&2874.7(3)&(3/2)$^-$&1653.8(3)&10.7(5)&0.019(1)\\
&&3951.2(3)&(3/2$^-$,5/2$^-$)&577.3(3)&28(1)&0.050(3)\\
4606.1(3)&(5/2$^+$)&0.0&7/2$^-$&4606(3)&100(3)&0.0113(7)\\
&&2578.1(3)&3/2$^+$&2028.0(4)&38(3)&0.0043(5)\\
&&3951.2(3)&(3/2$^-$,5/2$^-$)&653.2(4)&34(2)&0.0039(3)\\
4989.5(3)&3/2$^+$&0.0&7/2$^-$&4987(3)&0.9(6)&0.0011(7)\\
&&2013.4(3)&3/2$^-$&2976.2(4)&12.7(2)&0.0155(5)\\
&&2578.1(3)&3/2$^+$&2411.5(3)&71(1)&0.087(3)\\
&&2599.2(3)&1/2$^+$&2390.2(3)&20.0(4)&0.024(1)\\
&&2849.9(3)&(3/2)$^-$&2139.1(4)&2.4(1)&0.0030(2)\\
&&2874.7(3)&(3/2)$^-$&2114.6(4)&4.4(1)&0.0054(2)\\
&&3266.3(3)&5/2$^-$&1723.3(6)&0.34(4)&0.00042(5)\\
&&3951.2(3)&(3/2$^-$,5/2$^-$)&1038.4(4)&100(2)&0.123(3)\\
&&4606.1(3)&(5/2$^+$)&383.4(6)&3.9(2)&0.0048(3)\\
5305.9(3)&3/2$^+$&0.0&7/2$^-$&5304(4)&0.55(7)&0.00053(7)\\
&&2013.4(3)&3/2$^-$&3292(2)&74(1)&0.071(2)\\
&&2578.1(3)&3/2$^+$&2728.2(3)&100(2)&0.095(2)\\
&&2599.2(3)&1/2$^+$&2707.0(3)&40.0(7)&0.038(1)\\
&&2849.9(3)&(3/2)$^-$&2455.7(3)&12.3(3)&0.0118(4)\\
&&2874.7(3)&(3/2)$^-$&2431.2(3)&85(1)&0.081(2)\\
&&3266.3(3)&5/2$^-$&2039.4(4)&0.92(7)&0.00088(7)\\
&&3889.0(3)&(5/2$^-$)&1416.9(3)&9.5(3)&0.0091(4)\\
&&4606.1(3)&(5/2$^+$)&699.5(3)&8.2(7)&0.0078(7)\\
5456.1(3)&3/2$^+$&0.0&7/2$^-$&5454(4)&0.54(6)&0.0009(1)\\
&&2013.4(3)&3/2$^-$&3442.3(4)&0.14(5)&0.00024(8)\\
&&2578.1(3)&3/2$^+$&2878.1(3)&100(2)&0.171(6)\\
&&2599.2(3)&1/2$^+$&2856.8(3)&96(2)&0.163(4)\\
&&2849.9(3)&(3/2)$^-$&2605.7(3)&2.1(1)&0.0036(2)\\
&&2874.7(3)&(3/2)$^-$&2581.3(3)&18.4(5)&0.031(1)\\
&&3889.0(3)&(5/2$^-$)&1567.6(4)&0.54(7)&0.0009(1)\\
&&3951.2(3)&(3/2$^-$,5/2$^-$)&1505.1(3)&13.2(7)&0.023(1)\\
&&4528.3(3)&3/2$^+$&928.1(4)&1.3(2)&0.0023(3)\\
&&4606.1(3)&(5/2$^+$)&849.2(3)&4.7(5)&0.0081(8)\\
6064.2(4)&3/2$^+$&2599.2(3)&1/2$^+$&3464.7(4)&100(4)&0.0027(2)\\
&&2874.7(3)&(3/2)$^-$&3189.5(3)&92(4)&0.0025(1)\\
    \end{tabular}
    \end{ruledtabular}
\end{table*}
The relative intensities of the $\gamma$ rays from the $\beta^-$ decay of $^{47}$K were measured. The areas of peaks that could be identified in the energy spectrum were extracted using a Gaussian + skewed-Gaussian peak-fitting routine that included a background fit \cite{TPeak}. Peak areas were corrected for their relative efficiencies, as well as summing-in and summing-out effects (using the method described in Ref. \cite{Garnsworthy2019}) to determine the intensities.

For peaks that were not visible in the singles $\gamma$-ray energy spectrum, the process was more extensive. As an intermediate step, branching ratios for transitions from each state were extracted. If possible, a $\gamma$ ray that populated a state was identified and used as a gate. Each peak corresponding to a draining $\gamma$ ray of that state was fit to extract a peak area that was corrected for efficiency and summing. The ratios of these areas were converted into branching ratios for the draining transitions. If it was not possible to gate on a populating $\gamma$ ray, then branching ratios were extracted by gating on coincident $\gamma$ rays lower in the level scheme, and correcting the coincident peak area for summing and the energy-dependent efficiencies of both $\gamma$ rays, as well as the branching ratio of the gated transition. These corrected peak areas were then combined to calculate branching ratios for the higher-lying state. As at least one transition draining each state was visible in the energy spectrum, the intensity of that transition was used to scale the branching ratios of the other transitions and obtain the intensity relative to the 2013.1~keV $\gamma$ ray. The relative intensities, along with branching ratios of all $\gamma$ rays observed in this work are presented in Table \ref{table:intensities}.

\subsection{\label{ssec:bfeeding}\texorpdfstring{$\beta^-$}{beta}-decay branching ratios and log({\it ft}) values}

\begin{table*}[t]
    \caption{Energies, $\beta^-$ branching ratios, log $ft$ values, spins and parities, and $B(GT)$ values measured in this work and previous works. Asterisks indicate a log $ft$ value calculated for a unique, first-forbidden decay. All limits are given at the 95\% confidence level. Previously established energies, spins, and parities are from Ref. \cite{Burrows2007}. Literature $\beta^-$ branching ratios and $B(GT)$ values are from Ref. \cite{Alburger1984}.}
	\label{table:betafeeding}
	\begin{ruledtabular}
    \begin{tabular}{ccccc|ccccc}
    \multicolumn{5}{c|}{This work} & \multicolumn{5}{c}{Previous measurements}\\
	&&&&$B(GT)$&&&&&$B(GT)$\\
$E_s$ (keV)&$\beta^-$ branching&log $ft$&$J^\pi$&$(\times 10^{-3})$&$E_s$ (keV)&$\beta^-$ branching&log $ft$&$J^\pi$&$(\times 10^{-3})$\\
\hline
2013.4(3)&$<$2.9&$>$6.5&3/2$^-$&&2013.53(10)&$<$2.0&$>$6.68&3/2$^-$&\\
2578.1(3)&18.4(3)&5.457(8)&3/2$^+$&21.5(4)&2578.33(10)&19.0(3)&5.45(1)&3/2$^+$&21.7(4)\\
2599.2(3)&80(2)&4.807(9)&1/2$^+$&96(2)&2599.53(11)&81.0(15)&4.81(1)&1/2$^+$&95.1(18)\\
2849.9(3)&0.013(1)&8.46(4)&(3/2)$^-$&&2849(5)&&&(1/2,3/2)$^-$&\\
2874.7(3)&0.015(3)&8.40(9)&(3/2)$^-$&&2875.19(18)&&&(1/2,3/2)$^-$&\\
3266.3(3)&0.0038(2)&10.28(2)*&5/2$^-$&&3267(8)&&&(5/2,7/2)$^-$&\\
3889.0(3)&$<$0.0017&$>$10.1*&(5/2$^-$)&&3877(8)&&&(5/2)$^-$&\\
3951.2(3)&0.010(5)&7.9(2) or 9.3(2)*&(3/2$^-$,5/2$^-$)&&&&&&\\
4453.1(4)&0.036(5)&6.99(6)&(1/2,3/2)&&4455(8)&&&&\\
4528.3(3)&0.38(1)&5.91(1)&3/2$^+$&7.6(2)&4531(8)&&&(3/2)$^+$&\\
4606.1(3)&$<$0.0027&$>$8.0&(5/2$^+$)&&4611(8)&&&(5/2)$^+$&\\
4989.5(3)&0.246(6)&5.64(1)&3/2$^+$&14.1(3)&4988(8)&&&(3/2,5/2)$^+$&\\
5305.9(3)&0.291(7)&5.20(1)&3/2$^+$&38.9(9)&5305(5)&&&(3/2,5/2)$^+$&\\
5456.1(3)&0.38(1)&4.87(1)&3/2$^+$&83(2)&5459(5)&&&(3/2,5/2)$^+$&\\
6064.2(4)&0.0049(2)&5.56(2)&3/2$^+$&17.0(8)&6062(5)&&&(3/2,5/2)$^+$&\\
    \end{tabular}
    \end{ruledtabular}
\end{table*}

To determine the $\beta^-$-decay feeding to each of the observed states from the $1/2^+$ $^{47}$K ground state, two assumptions were made. The first was that feeding to the $7/2^-$ $^{47}$Ca ground state was negligible; if it occurred, this would be a third-forbidden transition. Grinyer, {\it et al.} \cite{Grinyer2005} estimated an unobserved decay to the ground state of $^{26}$Mg, a second-forbidden decay, as 0.01 ppm, which is 200 times smaller than the smallest $\beta^-$-decay branching ratio uncertainty calculated here. The second assumption was that contributions from weakly populated higher-lying states through unobserved transitions to lower states, the pandemonium effect \cite{Hardy1977}, were also negligible. This second assumption will be further discussed in Sec. \ref{ssec:bfeeding}. 

Given these assumptions, it is possible to determine the relative $\beta^-$-feeding without taking into account the $\beta^-$ detection of SCEPTAR. The relative intensities feeding and draining each state were tabulated. The difference between feeding and draining intensities was attributed to $\beta^-$ decay directly to the state.
Conversion coefficients for these transitions are expected to be orders of magnitude smaller than the relative uncertainties of these intensities and can be ignored.
Comparisons of each direct $\beta^-$-decay intensity to the total were made to provide the $\beta^-$-decay branching ratios given in Table \ref{table:betafeeding}.

The probability of direct $\beta^-$ decay to these states is influenced by the energy released in the decay as well as nuclear structure considerations, including angular momentum, that make a decay more or less likely. To eliminate the influence of that phase-space factor and gain insight into the nuclear structure influence on the decay, we calculate a log $ft$ value for each $\beta^-$-decay branch, using a $Q$-value of 6632.4(26)~keV and a half-life of 17.38(3) s (the value measured in this paper, see Sec. \ref{ssec:halflife}). For states that were determined to be of spin $1/2^+$ or $3/2^+$, a $B(GT)$ value was calculated where $B(GT)=6170/ft$ \cite{Alburger1984}. Both the log $ft$ values and $B(GT)$ values are listed in Table \ref{table:betafeeding}.


Previous $\beta^-$-decay studies of $^{47}$K only positively observed $\beta$-decay feeding to the lowest three excited states (see Table \ref{table:betafeeding}) and $\gamma$ rays with relative intensities down to 1.12\% (setting the intensity of the 2013.1~keV transition at 100\%) \cite{Alburger1984}. The calculated $\beta^-$ branching ratios were 81.0(15)\% and 19.0(3)\% branching to the first excited $1/2^+$ and $3/2^+$ states, respectively, via allowed transitions. As of the last data evaluation \cite{Burrows2007}, seventeen higher-lying states were candidates for population via allowed transitions as well. If all these states were populated and then each decayed to the $1/2^+$ and $3/2^+$ excited states with a relative intensity of 1\%, the $\beta^-$ branching ratios to the lowest $1/2^+$ and $3/2^+$ excited states would then be approximately 65\% and 3\%, far outside the present uncertainties.

The increased sensitivity of GRIFFIN allows the measurement of $\gamma$ rays with relative intensities below 0.001\% and expands the known level scheme to within 568(3)~keV of the 6.632~MeV $Q$-value. 
This increase in sensitivity allows a more rigorous constraint on the influence of the pandemonium effect on the branching ratios to the first three excited states.
Five of the seventeen higher-lying states that were candidates for population via allowed transitions have been included in the level scheme.
This leaves eleven states listed in Ref. \cite{Burrows2007} with tentative spins including $3/2^+$, and one state at 5785(8) keV has a spin of $1/2^+$, all of which would be allowed decays. If all 12 of these states were populated and decayed to all the observed states at lower energies, this would be 175 unobserved transitions. If we assume all of those transitions had relative intensities of 0.001\% (approximately our sensitivity limit), this would be a total missing relative intensity of 0.175\%. While this missing intensity could certainly impact the $\beta^-$-decay branching ratios to the states above 2600~keV, it is entirely within the uncertainties of the branching ratios to the states at 2578.1~keV and 2599.2~keV.


\subsection{\label{ssec:ang-cor}\texorpdfstring{$\gamma - \gamma$}{gamma-gamma} angular correlations}

\begin{figure}[t]
	\centering
    \includegraphics[width=0.95\linewidth]{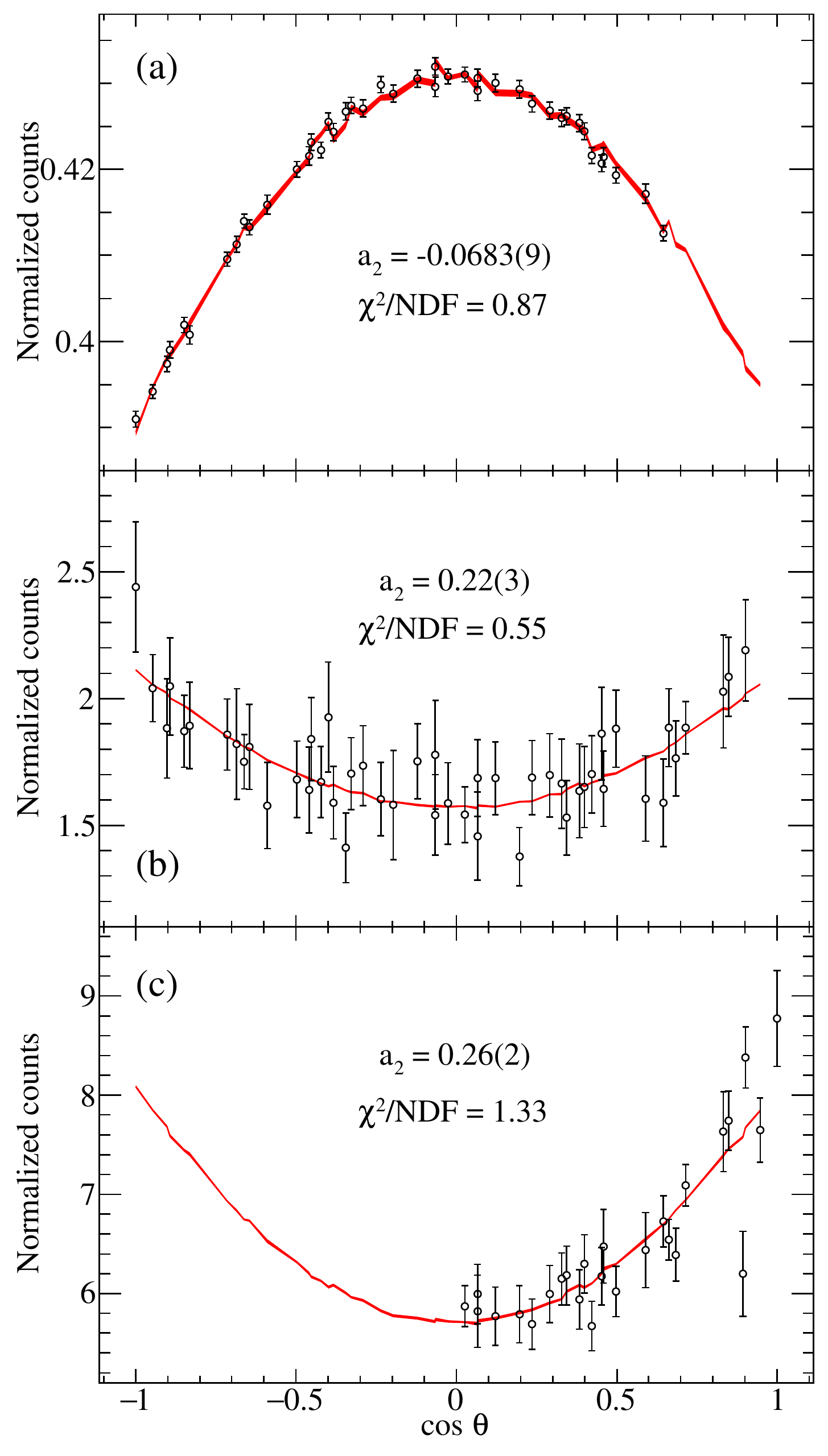}
    \caption{Angular correlations of three prominent $\gamma$-$\gamma$ cascades: (a) 586.1-2013.1~keV, (b) 2878.1-564.9~keV, and (c) 2411.5-564.9~keV. The data are shown as open circles with error bars, the best fit simulations and their uncertainties are shown with the red areas, and the minimized coefficients are presented for each cascade. For the cascade shown in panel (a), significant Compton background from other photopeaks interfered with proper peak fitting at small angles, so those data were excluded from the plots and the fits. For the cascade shown in panel (c), the data were folded about $\cos \theta=0$ to increase the statistics before fitting the peaks.}
  	\label{fig:gg-correlation-example}
\end{figure}

Experimental $\gamma$-$\gamma$ angular correlations were examined in order to assign or constrain the spins of excited states and determine the multipole mixing ratios of the $\gamma$ rays involved in the cascade.
A detailed description of the methods followed in this work are available in Ref. \cite{Smith2018}, which includes a treatment of the data using an event-mixing approach to account for variations in individual crystal efficiencies, and corrections for the effects of finite solid angle due to the large size of the crystals. For most of the cascades in this work, we followed the Method 2 approach outlined in Ref. \cite{Smith2018} where separate GEANT4 simulations for each of the components of the Legendre polynomials were performed for the specific $\gamma$-ray energies involved in each cascade. The experimental data were fitted with a linear combination of the simulation results to properly account for the finite solid angle and other experimental effects. 
Examples of the correlations for three cascades are shown in Fig. \ref{fig:gg-correlation-example}.
When the statistics were poorer, it became necessary to `fold' the angular correlation about an angular difference of 90$^{\circ}$. 
This technique combines energy spectra at angles equidistant from 90$^{\circ}$, taking advantage of the symmetry of all angular distributions about 90$^{\circ}$ to increase the statistics in a spectrum before fitting it with a peak shape. Folded angular correlations are then analyzed with folded simulations. An example is shown in Fig. \ref{fig:gg-correlation-example}(c). All cascades that were analyzed after folding the data are marked in Table \ref{table:angcorr} with a $^{\dagger}$.
For a handful of cascades with lower statistics, indicated in Table \ref{table:angcorr} with an asterisk, we used the Method 4 approach, where coefficients of a Legendre polynomial fit to the experimental spectrum were converted to values of $a_2$ and $a_4$ using energy-dependent scaling factors derived from simulation.

The results obtained for the $\gamma$-$\gamma$ angular correlations evaluated in $^{47}$Ca are presented as $a_2$ and $a_4$ values in Table \ref{table:angcorr}. Legendre coefficients $a_2$ and (if applicable) $a_4$ were measured from all correlations. The use of $\gamma$-$\gamma$ angular correlations to determine spin can be more challenging in an odd-$A$ nucleus than in an even-$A$ nucleus, because often cascades will have two potentially nonzero mixing ratios that need to be determined simultaneously. In contrast, even-$A$ nuclei typically can involve a state with $J=0$, which constrains the mixing ratio of at least one transition to be $\delta=0$. For some of the cascades in $^{47}$Ca, however, we were able to measure or constrain sufficiently one of the mixing ratios (always for the second $\gamma$ ray, noted in Table \ref{table:angcorr} as $\delta_{\gamma 2}$), which allowed us to measure or constrain the other mixing ratio (always for the first $\gamma$ ray, noted in Table \ref{table:angcorr} as $\delta_{\gamma 1}$). All of the cascades fit in this way were well fit with the cascade, defined as having a solution that was within a 99\% confidence limit \cite{Smith2018}. One $\gamma$ ray, 2976.2~keV, had a mixing ratio that was unconstrained by the angular correlation (2976.2-2013.1~keV).

\begin{table*}[t]
    \caption{Angular correlation results for $\gamma$-$\gamma$ cascades in $^{47}$Ca, organized by initial level energy ($E_{i}$) and identified by the energies of the first and second transitions ($E_{\gamma 1}$ and $E_{\gamma 2}$). All cascades were fit to extract Legendre coefficients $a_2$ and $a_4$. The best fit parameters ($a_2$ and $a_4$) and reduced $\chi^2$ value ($\chi^2/\nu$) are presented. Dashed entries indicate that the cascade cannot include a fourth-order component by angular momentum considerations. Cascades where the mixing ratio for the second $\gamma$ ray ($\delta_{\gamma 2}$) could be deduced were also fit to extract a mixing ratio for the first $\gamma$ ray ($\delta_{\gamma 1}$), assuming a cascade of the presented spins ($J_{i} \rightarrow J_{m} \rightarrow J_{f}$). Most cascades were fit with Method 2 of Ref. \cite{Smith2018}; asterisks indicate that the fit was performed with Method 4. Cascades that were folded are indicated with a $^{\dagger}$.}
	\label{table:angcorr}
	\begin{ruledtabular}
    \begin{tabular}{ccc|ccc|ccc}
    $E_{i}$ & $E_{\gamma_1}$ & $E_{\gamma_2}$ & $a_2$ & $a_4$ & $\chi^2/\nu$ & $J_{i} \rightarrow J_{m} \rightarrow J_{f}$ & $\delta_{\gamma2}$ & $\delta_{\gamma1}$\\
    \hline
	2578.1 & 564.9	& 	2013.1	&	0.059(2)		& 	-	& 	1.07 & $3/2\rightarrow3/2\rightarrow7/2$ & 0.0000(38) & -0.007(11)\\
	2599.2 & 586.1	& 	2013.1	&	-0.0683(9)		&  -	& 	0.87 & $1/2\rightarrow3/2\rightarrow7/2$ & 0.0000(38) & -0.013(9)\\
	2874.4 & 861.5	& 	2013.1	&	-0.02(2)		& 	-	& 	1.09 & $3/2\rightarrow3/2\rightarrow7/2$ & 0.0000(38) & 0.4(1) or 10(+200,-5)\\
	$^{\dagger}$3951.2 & 1373.1	& 	564.9	&	-0.05(3)		& 	-	& 	0.86  & $3/2\rightarrow3/2\rightarrow3/2$ & -0.007(11) & 0.36(+7,-6) or 10(+17,-4)\\
	& & & & & & $5/2\rightarrow3/2\rightarrow3/2$ & -0.007(11) & -0.03(7) or 5(+3,-1)\\
	4528.3 & 2515.0	& 	2013.1	&	0.09(3)		& 	-	& 	0.60 & $3/2\rightarrow3/2\rightarrow7/2$ & 0.0000(38) & -2.3(+8,-13) or -0.2(+1,-2)\\
	*$^{\dagger}$4528.3  & 1653.8	& 	861.5	&	0.04(5)		& 	-	& 1.15	& & &\\
	*$^{\dagger}$4528.3 & 577.3	& 	3951	&	0.01(5)		& 	0.12(8)	& 0.96	& & &\\
	$^{\dagger}$4989.5 & 2976.2	& 	2013.1	&	-0.01(9)		& 	-	& 0.76 & $3/2\rightarrow3/2\rightarrow7/2$ & 0.0000(38) & unconstrained\\
	$^{\dagger}$4989.5 & 2411.5	& 	564.9	&	0.26(2)		& 	-	& 	1.33 & $3/2\rightarrow3/2\rightarrow3/2$ & -0.007(11) & -2.2(+2,-3) or -0.17(4)\\
	*$^{\dagger}$4989.5 & 1038.4	& 	3951	&	0.04(4)		& 	-0.04(6)	& 0.29	& & &\\
	$^{\dagger}$4989.5 & 1038.4	& 	1373.1	&	0.16(2)		& 	-0.02(3)	& 1.58	& & &\\
	*5305.9 & 3292	& 	2013.1	&	0.09(3)		& 	-	& 0.84 & $3/2\rightarrow3/2\rightarrow7/2$ & 0.0000(38) & -2.3(+8,-14) or -0.2(+1,-2)	\\
	5305.9 & 2728.2	& 	564.9	&	0.21(7)		& 	-	&  0.34	& $3/2\rightarrow3/2\rightarrow3/2$ & -0.007(11) & -2.9(+8,-15) or -0.1(1)\\
	$^{\dagger}$5305.9 & 2431.2	& 	861.5	&	0.04(3)		& 	-	& 0.89 & & &	\\
	5456.1 & 2878.1	& 	564.9	&	0.22(3)		& 	-	& 	0.55 & $3/2\rightarrow3/2\rightarrow3/2$ & -0.007(11) & -2.8(+4,-6) or -0.09(6)\\
	*${\dagger}$5456.1 & 2581.3	& 	861.5	&	0.05(4)		& 	-	& 1.09 & & &	\\
    \end{tabular}
    \end{ruledtabular}
\end{table*}

\subsection{\label{ssec:halflife}Half-life of \texorpdfstring{$^{47}$K}{47K}}
The longer cycles with a source decay time of 350 seconds were used to measure the half-life of $^{47}$K, previously established as 17.50(24)~s \cite{Burrows2007} from a weighted average of 17.5(4) s and 17.5(3) s. 73 of these longer cycles were included in the half-life measurement. Data from an additional sixteen crystals impacted by electronics issues were excluded, leaving data from 40 of the 60 crystals. A channel-by-channel dead-time correction was applied to the data using a dead time of 6 $\mu$s per event.
\begin{figure}
    \centering
   \includegraphics[width=0.95\linewidth]{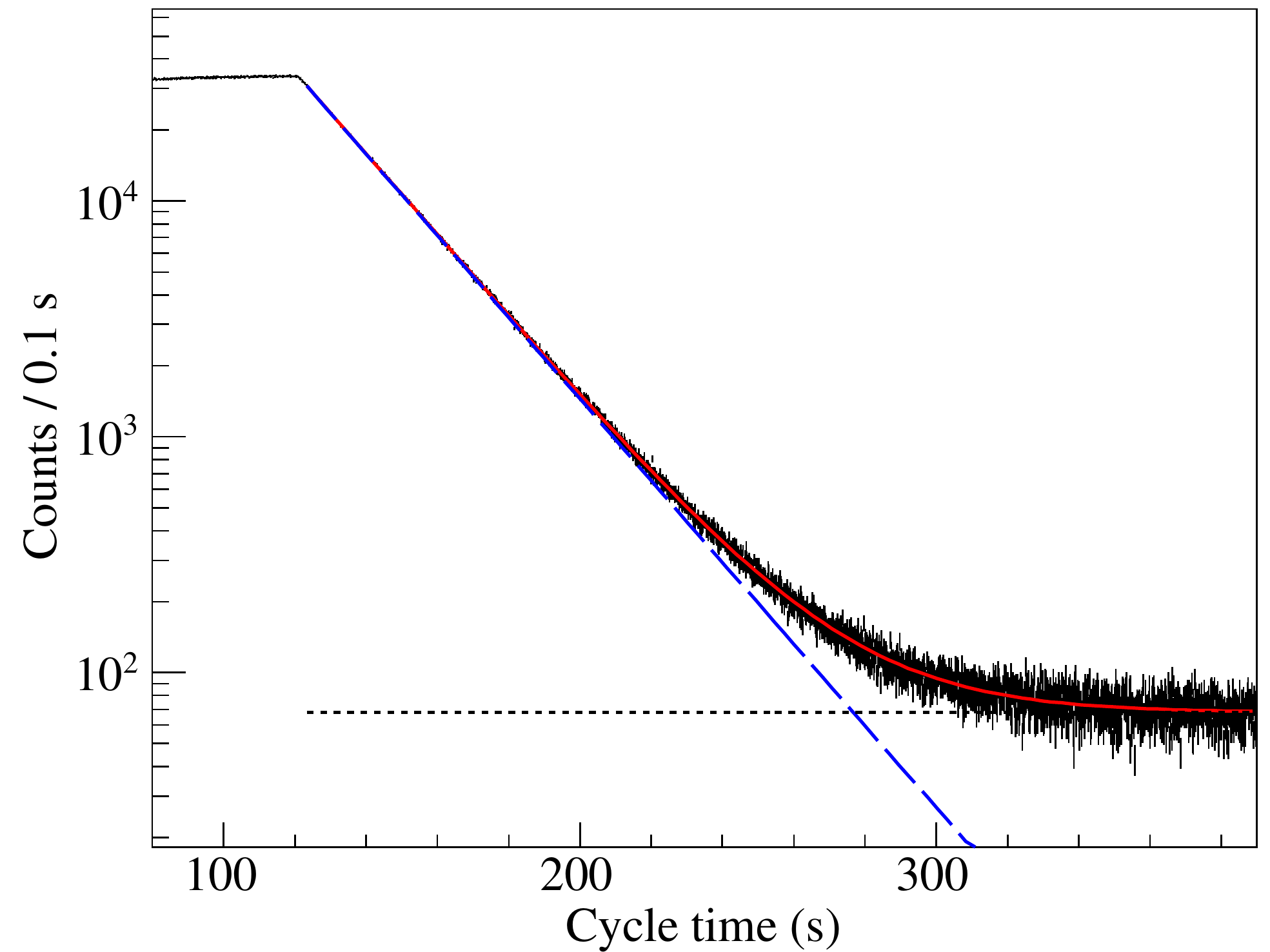}
    \caption{\label{fig:half-life}The decay of $^{47}$K based on the 586.1~keV $\gamma$ ray in $^{47}$Ca. The beam shut off at a cycle time of 121 s, so the fit begins at a time of 122 s. The exponential decay component of the fit is shown with the blue dashed line, the constant background component of the fit is shown with the black dotted line, and the total fit is shown with the red solid line.}
\end{figure}

The time distribution of all events associated with the three most intense $\gamma$ rays (564.9~keV, 586.1~keV, and 2013.1~keV) were separately fitted with a sum of an exponential function and a constant background.
The beam shut off at a cycle time of 121 s, and the fit included times from 122 to 400 s, or approximately 16 half-lives. The time distribution of the 586.1~keV $\gamma$ ray and its fit are shown in Fig. \ref{fig:half-life}. Best fits were $t_{1/2}=17.37(2)$ s, 17.380(7) s, and 17.385(8) s for the 564.9, 586.1, and 2013.1~keV $\gamma$ rays, respectively.

A systematic manipulation of the binning of the decay curves revealed no systematic deviation from the measured half-life. Small variations of the energy gate limits showed some deviation in the best-fit half-lives, and additional uncertainties of 0.02 s, 0.01 s, and 0.02 s were assigned to the three individual half-life measurements. A systematic change of the fit range showed some dependence on the selection of the first bin. Systematic uncertainties of 0.05 s, 0.04 s, and 0.07 s were assigned for this. Combining statistical and systematic uncertainties, the measured half-life with the 564.9~keV, 586.1~keV, and 2013.1~keV $\gamma$ rays were 17.37(6) s, 17.38(4) s, and 17.38(7) s. A weighted average of these three gives a final half-life value from this work of 17.38(3) s which is in agreement with, but a factor of 8 more precise than the previous evaluated value.

\section{\label{sec:discussion}Discussion}
A more complete understanding of the structure of this nucleus with one neutron hole can be obtained by combining all the information now available from previous studies of $^{47}$Ca, with the log $ft$ values, angular correlation coefficients, and $\gamma$-ray branching ratios measured in this work. The results are presented in Tables \ref{table:intensities} and \ref{table:betafeeding}. Discussion of the conclusions for each excited state are presented here. The states are presented in an order appropriate for building the conclusions, not necessarily by energy.


\subsection{\label{ssec:2013}2013.4~keV state}
The spin of the first excited state at 2013.4~keV is 3/2$^-$ and has a single draining transition to the 7/2$^-$ ground state. From spin and parity considerations, the transition will be of $E2/M3$ character. With this combination, it is also likely to be a fairly pure transition; Ref. \cite{Montanari2012} measured the mixing ratio as -0.17, albeit from a single data point and without any error bars. In addition, the lifetime of the state was measured as 6.1(3) ps.

The mixing ratio of the 2013.1~keV $\gamma$ ray can be constrained with a recommended upper limit for reduced transition rates derived from a systematic study of this mass region \cite{Endt1979}. Assuming the recommended upper limit of $B(M3)\downarrow=10$ W.u., the magnitude of the mixing ratio of the 2013.1~keV $\gamma$ ray must be less than 0.0038. This range is consistent with all the angular correlations made with this $\gamma$ ray.
%
%
For the rest of the analysis, we will take the value as $\delta_{2013.1}$=0.0000(38). With that, the $B(E2)\downarrow$ for the 2013.1~keV $\gamma$ ray to the ground state would be $B(E2)\downarrow=0.40(2)$ W.u..

\subsection{\label{ssec:2577}2578.1~keV state}
The spin of the second excited state has been established as $3/2^+$ in a transfer reaction study \cite{Fortier1978}. The lifetime of this state has been measured as 9.6(14) ps \cite{Montanari2012}. The decay to this state is therefore an allowed Gamow-Teller transition from the $1/2^+$ ground state of $^{47}$K. The log $ft$ value of 5.457(8) is consistent with this spin assignment (Table \ref{table:betafeeding}). The angular correlation of the 564.9-2013.1~keV cascade is also consistent with the spin and parity assignment of $3/2^+$.

The mixing ratio for the 564.9~keV $\gamma$ ray can be constrained by both the angular correlation of 564.9-2013.1~keV cascade and the recommended upper limit for an $M2$ transition.
From the angular correlation, and assuming $\delta_{2013.1}$=0.0000(38), the value of the 564.9~keV mixing ratio can be either $\delta_{564.9}=-0.007(11)$ or $\delta_{564.9}=-3.8(1)$.
From a recommended upper limit of $B(M2)\downarrow$ = 1 W.u. \cite{Endt1979}, the magnitude of the mixing ratio should be less than 0.015, so we choose the former option and set $\delta_{564.9}=-0.007(11)$ for the remaining analysis. This choice is consistent with the other four angular correlations that involve the 564.9~keV $\gamma$ ray. The associated $E1$ reduced transition rate is $B(E1)\downarrow=0.00031(+5,-4)$ W.u..


With the branching ratio measurement for the 2578.1~keV $\gamma$ ray presented in this work, a pure $M2$ transition would have a $B(M2)\downarrow$ of 0.9(2) W.u.. Assuming an upper limit of 100 W.u. for the limit on $B(E3)\downarrow$ \cite{Endt1979}, this would limit the mixing ratio of the 2578.1~keV $\gamma$ ray to less than 0.47. A lower recommended upper limit of 50 W.u., which might be suggested by a recent reduction in the $B(E3)$ recommended upper limit in the $A=4-44$ region \cite{Endt1993}, would limit the mixing ratio of the 2578.1~keV $\gamma$ ray to less than 0.20.

\subsection{\label{ssec:2598}2599.2~keV state}
As expected, the majority of the $\beta^-$-decay strength goes directly to the state at 2599.2~keV (Table \ref{table:betafeeding}) that is known to have a spin-parity of $1/2^+$ \cite{Fortier1978} and a lifetime of 6.8(6) ps \cite{Montanari2012}. The log $ft$ value measured here of 4.807(9) is consistent with that of an allowed decay.

Just as in the previous section, the mixing ratio of the 586.1~keV transition is constrained by angular correlation measurements and the recommended upper limits on reduced transition rates. The mixing ratios consistent with the angular correlation of the 586.1-2013.1~keV cascade are $\delta_{586.1}=-0.013(9)$ and $\delta_{586.1}=1.78(+4,-2)$. The 1 W.u. recommended upper limit for an $M2$ transition limits the mixing ratio to $|\delta_{586.1}|<0.012$ though, so the larger range can be excluded, leaving the mixing ratio as $\delta_{586.1}=-0.013(9)$ and thus $B(E1)\downarrow=0.00054(+5,-4)$ W.u..

Using the branching ratio measurements presented here, if the 2598.5~keV $\gamma$ ray were a pure $E3$ transition, it would have a reduced transition probability of $B(E3)\downarrow=11(1)$~W.u.. The recommended upper limit of 30 W.u. for $M4$ transitions \cite{Endt1979} sets an upper limit of 0.0014 on the mixing ratio of the 2598.5~keV $\gamma$ ray.

\subsection{\label{ssec:highestlevels}States above 4700~keV}
Above 4700~keV, there are four states seen in this work that have all been previously assigned tentative spin-parities of $(3/2,5/2)^+$ from transfer reaction measurements \cite{Belote1966,Fortier1978}. The log $ft$ values for all of these states range between 4.87(1) and 5.64(1) and are all consistent with allowed decays, eliminating $5/2^+$ as a spin-parity possibility. Low statistics prevent $\gamma$-$\gamma$ angular correlations associated with the 6064.2~keV state to be constructed, but $\gamma$-$\gamma$ angular correlations of $\gamma$ rays depopulating the 4989.5, 5305.9, and 5456.1~keV states are possible with the 861.5, 564.9, and 2013.1~keV $\gamma$ rays, respectively. All of these cascades are consistent with a $3/2$ spin assignment, and those that include the 2013.1 or 564.9~keV $\gamma$ rays are consistent with the mixing ratio constraints established earlier. The angular correlations provided constraints on the mixing ratios for four $\gamma$ rays directly emitted from these states (Table \ref{table:angcorr}).
Based on this information, we assign all of these states as $3/2^+$.

With this spin assignment, we can follow a similar procedure to that used in Ref. \cite{Alburger1984} and calculate $B(GT)$ values to compare the strength of $1s_{1/2}$ proton hole configurations in each state (Table \ref{table:betafeeding}). These $B(GT)$ values range from $14.1(3)\times10^{-3}$ to $83(2)\times10^{-3}$, indicating strengths comparable to those of the much more strongly populated $3/2^+$ and $1/2^+$ states at 2578.1 and 2599.2~keV, respectively. Combined with the 4528.3~keV state (Sec. \ref{ssec:4527}), these states are now found to comprise a total $\beta^-$-feeding intensity of 1.29(2)\%, over half of the observed $1s_{1/2}$ neutron and proton hole strengths in $^{47}$Ca and 88(1)\% for $1s_{1/2}$ proton hole states of the type $[(\pi 1s_{1/2})^{-1}(\nu 0f,1p)^7(\pi 0f,1p)]$ coupled to $3/2^+$. The finding of significant strength at this energy is consistent with the structure interpretation of similar excited states in $^{48,49}$Ca \cite{Multhauf1975,Huck1981,Carraz1982}.

\subsection{\label{ssec:2874}2849.9~keV and 2874.7~keV states}
The possible spins of the states at 2849.9~keV and 2874.7~keV have been constrained by transfer measurements to be of $L=1$ character, so $J=(1/2,3/2)^-$ \cite{Burrows2007}. $\beta^-$ decay to either spin would be classified as a non-unique, first-forbidden transition, so the log $ft$ values cannot be used to discriminate, though at 8.46(4) and 8.40(9), both are consistent with first-forbidden decay.

No angular correlations involving the 2849.9~keV state were possible.
Four different angular correlations involving the 861.5~keV $\gamma$ ray from the 2874.7~keV state were constructed (Table \ref{table:angcorr}), three of which use the 2874.7~keV state as an intermediate state. Using the limits of $\delta_{2013.1}$ found earlier, a mixing ratio analysis can be performed on the 861.5-2013.1~keV cascade. With such an analysis, both $3/2^-$ and $1/2^-$ spin assignments are consistent with the data of the 861.5-2013.1~keV cascade, with neither being more strongly favored.

If the spin of the 2874.7~keV state is $1/2^-$, then the three angular correlations with it as an intermediate state must be isotropic; if the spin of the state is $3/2^-$, the angular correlations can only have one non-zero coefficient - $a_2$. All three of those angular correlations are consistent with isotropic distributions, so the $1/2^-$ spin cannot be eliminated. Unfortunately, an isotropic distribution is also consistent with a large range of mixing ratios associated with a $3/2^-$ spin assignment as well.

Significant differences between these states occur in the branching ratios for their draining transitions. The 2849.9~keV state only had a single visible transition to the $7/2^-$ ground state, while the 2874.7~keV state decays principally to the $3/2^-$ 2013.4~keV state. A state with a spin of 3/2$^-$ would decay to states of $7/2^-$ and $3/2^-$ with $E2$ and $M1$ transitions, respectively, while a state with a spin of 1/2$^-$ would decay to those states with $M3$ and $M1$ transitions. The observed decay strengths then, suggest a 3/2$^-$ spin for both states, though we leave them both as tentative.


\subsection{\label{ssec:4527}4528.3~keV state}
Transfer reaction measurements \cite{Fortier1978,Hanspal1985} identified a state at 4531(8) keV with a tentative spin-parity of $3/2^+$ based on a coupled reaction channel analysis. Such a state would be populated by an allowed decay transition in the $\beta^-$ decay of $^{47}$K. The log $ft$ value is measured as 5.91(1), which is consistent with an allowed decay. The strongest $\gamma$-$\gamma$ cascade coming from this state, 2515.0-2013.1~keV is consistent with the mixing ratio constraints placed earlier on the 2013.1~keV $\gamma$ ray and a spin of $3/2$. It also provides a constraint on the mixing ratio of the 2515.0~keV $\gamma$ ray (Table \ref{table:angcorr}). Based on this information, the 4528.3~keV state is confirmed as $J^\pi=3/2^+$. The associated $B(GT)$ value for this state is $7.6(2)\times10^{-3}$, the smallest value observed in this study.

\subsection{\label{ssec:3950}3951.2~keV state}
No state near this energy has been seen during previous experiments, so any constraints placed on the spin depend entirely on the data collected in this work.

Assuming an allowed or non-unique decay, the $\beta^-$ decay to this state has a log $ft$ value of 7.9(2), which could be either an allowed or first-forbidden transition and would be consistent with spins of $1/2^+,3/2^+,1/2^-$, or $3/2^-$. Calculated as a unique, first-forbidden decay, the log $ft$ value would be 9.3(2), which would be consistent with a spin of $5/2^-$. The potential spin-parity values based on the log $ft$ value are therefore reasonably constrained as $1/2^+$, $3/2^+$, $1/2^-$, $3/2^-$, or $5/2^-$.

Four $\gamma$-$\gamma$ angular correlations were constructed that involve this state.
Three $\gamma$-$\gamma$ angular correlations were constructed that include this state as the intermediate state: 1038.4-1373.1~keV, 1038.4-3951~keV, and 577.3-3951~keV, all presented in Table \ref{table:angcorr}. The strong $a_2$ value for the 1038.4-1373.1~keV cascade ($a_2$=0.16(2)) eliminates a spin of $1/2$, since cascades with an intermediate state spin of $1/2$ are isotropic. 
The mixing ratio analysis of the 1373.1-564.9~keV cascade places different constraints on the mixing ratio of the 1373.1~keV $\gamma$ ray, depending on the spin of the initial state (Table \ref{table:angcorr}).

Notably, this state has a strong decay branch to the ground state ($J^{\pi}=7/2^-$) and the first $3/2^+$ state, a weaker branch to the first $3/2^-$ excited state and no detected branch to the first $1/2^+$ excited state. The decay to the ground state would be a significantly enhanced $M2$ transition if the state had a spin of $3/2^+$, so we eliminate that possibility. A suppression of the decay to the $1/2^+$ excited state is expected if the spin were $5/2^-$ since this would be an $M2/E3$ transition.

Given the information here, we tentatively assign the spin and parity of this state to ($3/2^-$, $5/2^-$).

\subsection{\label{ssec:3265}3266.3~keV state}
A state at 3267(8) keV has been previously identified with a spin-parity constrained to $(5/2,7/2)^-$ by transfer reactions \cite{Fortier1978,Hanspal1985}.

$\beta^-$ decay to a $7/2^-$ state would be a non-unique, third-forbidden decay. Calculating the log $ft$ for a non-unique transition gives a value of 8.79(2), which is inconsistent with a third-forbidden decay. A decay to a $5/2^-$ state, however, would be a unique, first-forbidden transition. Calculated as a unique, first-forbidden decay, the log $ft$ value is 10.28(3), which is consistent. Combining this with the information from transfer measurements allows us to assign a spin-parity of $5/2^-$.

\subsection{\label{ssec:3888}3889.0~keV state}

No state exists in the literature with an excitation energy consistent with the energy measured here. The closest candidate is a state at 3877(8) keV that was identified in Ref. \cite{Fortier1978} and was observed to have non-pickup behavior. It was tentatively assigned a spin-parity of $(5/2^-)$ based on a coupled reaction channel analysis.

Assuming an allowed or non-unique transition, the log $ft$ value of this transition is $>$8.7, indicating that decay to this state is not an allowed transition.
If it is a unique, first-forbidden transition, the log $ft$ value of the $\beta^-$ decay is $>$10.1, consistent with a unique, first-forbidden transition. Assuming a unique, second-forbidden transition, the log $ft$ value would be $>$11.5, which is also consistent.
If the observed state is the same as the state identified in Ref. \cite{Fortier1978}, then there are no data that refute the tentative assignment of $(5/2^-)$.
 
\subsection{\label{ssec:4451}4453.1~keV state}
A state at 4455(8) was seen in a $^{48}$Ca$(^3$He,$\alpha)$ reaction measurement, but was not assigned a spin-parity \cite{Fortier1978,Hanspal1985}. In this work, the log $ft$ value was measured at 6.99(6), indicating an allowed or first forbidden decay. If we re-calculate the log $ft$ for unique first- and second-forbidden decays, we get 8.14(6) and 9.39(6), respectively. The former is consistent, but the latter is not.
Based on the log $ft$ value, this state could have a spin and parity of $1/2^+$, $3/2^+$, $1/2^-,3/2^-$, or $5/2^-$.
It was not possible to construct any angular correlations involving this state.

The observed decays from this state go to states with spins of $3/2^-$, $3/2^+$, and $3/2^-$. Notably, no transition to the $7/2^-$ ground state is observed, favoring a lower spin. Based on these observations, we tentatively assign this state a spin of (1/2,3/2), with no parity constraint.

\subsection{\label{ssec:4605}4606.1~keV state}
A state at 4611(8)~keV was previously observed \cite{Fortier1978,Hanspal1985}, which is consistent with the measured energy of 4606.1(3)~keV in this work. Based on a coupled reaction channel analysis, it was assigned a tentative spin-parity of $5/2^+$. The $\beta^-$-decay branching ratio to this state is consistent with zero and has a log $ft$ value of greater than 8.0 (95\% confidence limit), assuming an allowed or non-unique decay, or $>$9.1 or $>$10.3 assuming a unique first- or second-forbidden decay, respectively. This observation is consistent with the previous tentative assignment of 5/2$^+$, but this spin could not be confirmed independently in this work. Therefore, the spin assignment remains tentatively $(5/2^+)$.

\section{\label{sec:conclusion}Conclusion}
A detailed spectroscopic examination of the $\beta^-$ decay of $^{47}$K to $^{47}$Ca with the GRIFFIN spectrometer has led to the placement of 48 $\gamma$-ray transitions and twelve states to be firmly added to the $\beta^-$-decay level scheme. Of those twelve states, it was possible to newly assign or constrain the spins for eight of them.

The $\gamma$-ray detection in this study is three orders of magnitude more sensitive than the previous studies and reduces the gap between the $Q$-value and the highest known state in the decay scheme from approximately 4~MeV to 568(3)~keV, but the $\beta^-$-decay branching ratios determined here remain consistent with previous measurements. Five previously unobserved 3/2$^+$ states between $\sim$4.5 and 6.1~MeV excitation in $^{47}$Ca were observed with a total $\beta^-$-feeding intensity of 1.29(2)\%.  The sum of the $B(GT)$ values for these states is a measure of significant $1s_{1/2}$ proton hole strength near 5~MeV excitation, consistent with systematics of the neighboring $^{48,49}$Ca.

\begin{acknowledgments}
We would like to thank the operations and beam delivery staff at TRIUMF for providing the radioactive beam. J.K.S. would like to thank S.R. Stroberg, J. Henderson, J. Smallcombe, and J.R. Leslie for helpful discussions. The GRIFFIN spectrometer was jointly funded by the Canadian Foundation for Innovation (CFI), TRIUMF, and the University of Guelph. TRIUMF receives federal funding via a contribution agreement through the National Research Council Canada (NRC). C.E.S. acknowledges support from the Canada Research Chairs program. This work was supported in part by the Natural Sciences and Engineering Research Council of Canada (NSERC). This material is based upon work supported by the U.S. National Science Foundation (NSF) under Grant No. PHY-1913028.
\end{acknowledgments}

\bibliographystyle{apsrev}
\bibliography{Ca47paper}

\begin{thebibliography}{45}
\expandafter\ifx\csname natexlab\endcsname\relax\def\natexlab#1{#1}\fi
\expandafter\ifx\csname bibnamefont\endcsname\relax
  \def\bibnamefont#1{#1}\fi
\expandafter\ifx\csname bibfnamefont\endcsname\relax
  \def\bibfnamefont#1{#1}\fi
\expandafter\ifx\csname citenamefont\endcsname\relax
  \def\citenamefont#1{#1}\fi
\expandafter\ifx\csname url\endcsname\relax
  \def\url#1{\texttt{#1}}\fi
\expandafter\ifx\csname urlprefix\endcsname\relax\def\urlprefix{URL }\fi
\providecommand{\bibinfo}[2]{#2}
\providecommand{\eprint}[2][]{\url{#2}}

\bibitem[{\citenamefont{Hagen et~al.}(2012)\citenamefont{Hagen, Hjorth-Jensen,
  Jansen, Machleidt, and Papenbrock}}]{Hagen2012}
\bibinfo{author}{\bibfnamefont{G.}~\bibnamefont{Hagen}},
  \bibinfo{author}{\bibfnamefont{M.}~\bibnamefont{Hjorth-Jensen}},
  \bibinfo{author}{\bibfnamefont{G.~R.} \bibnamefont{Jansen}},
  \bibinfo{author}{\bibfnamefont{R.}~\bibnamefont{Machleidt}},
  \bibnamefont{and}
  \bibinfo{author}{\bibfnamefont{T.}~\bibnamefont{Papenbrock}},
  \bibinfo{journal}{Phys. Rev. Lett.} \textbf{\bibinfo{volume}{109}},
  \bibinfo{pages}{032502} (\bibinfo{year}{2012}), ISSN
  \bibinfo{issn}{0031-9007},
  \urlprefix\url{http://link.aps.org/doi/10.1103/PhysRevLett.109.032502}.

\bibitem[{\citenamefont{Holt et~al.}(2012)\citenamefont{Holt, Otsuka, Schwenk,
  and Suzuki}}]{Holt2012}
\bibinfo{author}{\bibfnamefont{J.~D.} \bibnamefont{Holt}},
  \bibinfo{author}{\bibfnamefont{T.}~\bibnamefont{Otsuka}},
  \bibinfo{author}{\bibfnamefont{A.}~\bibnamefont{Schwenk}}, \bibnamefont{and}
  \bibinfo{author}{\bibfnamefont{T.}~\bibnamefont{Suzuki}},
  \bibinfo{journal}{J. Phys. G Nucl. Part. Phys.}
  \textbf{\bibinfo{volume}{39}}, \bibinfo{pages}{085111}
  (\bibinfo{year}{2012}), ISSN \bibinfo{issn}{0954-3899},
  \urlprefix\url{http://stacks.iop.org/0954-3899/39/i=8/a=085111}.

\bibitem[{\citenamefont{Hergert et~al.}(2013)\citenamefont{Hergert, Bogner,
  Binder, Calci, Langhammer, Roth, and Schwenk}}]{Hergert2013}
\bibinfo{author}{\bibfnamefont{H.}~\bibnamefont{Hergert}},
  \bibinfo{author}{\bibfnamefont{S.~K.} \bibnamefont{Bogner}},
  \bibinfo{author}{\bibfnamefont{S.}~\bibnamefont{Binder}},
  \bibinfo{author}{\bibfnamefont{A.}~\bibnamefont{Calci}},
  \bibinfo{author}{\bibfnamefont{J.}~\bibnamefont{Langhammer}},
  \bibinfo{author}{\bibfnamefont{R.}~\bibnamefont{Roth}}, \bibnamefont{and}
  \bibinfo{author}{\bibfnamefont{A.}~\bibnamefont{Schwenk}},
  \bibinfo{journal}{Phys. Rev. C} \textbf{\bibinfo{volume}{87}},
  \bibinfo{pages}{034307} (\bibinfo{year}{2013}), ISSN
  \bibinfo{issn}{0556-2813},
  \urlprefix\url{http://link.aps.org/doi/10.1103/PhysRevC.87.034307}.

\bibitem[{\citenamefont{Som{\`{a}} et~al.}(2014)\citenamefont{Som{\`{a}},
  Cipollone, Barbieri, Navr{\'{a}}til, and Duguet}}]{Soma2014}
\bibinfo{author}{\bibfnamefont{V.}~\bibnamefont{Som{\`{a}}}},
  \bibinfo{author}{\bibfnamefont{A.}~\bibnamefont{Cipollone}},
  \bibinfo{author}{\bibfnamefont{C.}~\bibnamefont{Barbieri}},
  \bibinfo{author}{\bibfnamefont{P.}~\bibnamefont{Navr{\'{a}}til}},
  \bibnamefont{and} \bibinfo{author}{\bibfnamefont{T.}~\bibnamefont{Duguet}},
  \bibinfo{journal}{Phys. Rev. C} \textbf{\bibinfo{volume}{89}},
  \bibinfo{pages}{061301(R)} (\bibinfo{year}{2014}), ISSN
  \bibinfo{issn}{0556-2813},
  \urlprefix\url{http://link.aps.org/doi/10.1103/PhysRevC.89.061301}.

\bibitem[{\citenamefont{Holt et~al.}(2014)\citenamefont{Holt, Men{\'{e}}ndez,
  Simonis, and Schwenk}}]{Holt2014}
\bibinfo{author}{\bibfnamefont{J.~D.} \bibnamefont{Holt}},
  \bibinfo{author}{\bibfnamefont{J.}~\bibnamefont{Men{\'{e}}ndez}},
  \bibinfo{author}{\bibfnamefont{J.}~\bibnamefont{Simonis}}, \bibnamefont{and}
  \bibinfo{author}{\bibfnamefont{A.}~\bibnamefont{Schwenk}},
  \bibinfo{journal}{Phys. Rev. C} \textbf{\bibinfo{volume}{90}},
  \bibinfo{pages}{024312} (\bibinfo{year}{2014}), ISSN
  \bibinfo{issn}{0556-2813},
  \urlprefix\url{http://link.aps.org/doi/10.1103/PhysRevC.90.024312}.

\bibitem[{\citenamefont{Jansen et~al.}(2016)\citenamefont{Jansen, Schuster,
  Signoracci, Hagen, and Navratil}}]{Jansen2016}
\bibinfo{author}{\bibfnamefont{G.~R.} \bibnamefont{Jansen}},
  \bibinfo{author}{\bibfnamefont{M.~D.} \bibnamefont{Schuster}},
  \bibinfo{author}{\bibfnamefont{A.}~\bibnamefont{Signoracci}},
  \bibinfo{author}{\bibfnamefont{G.}~\bibnamefont{Hagen}}, \bibnamefont{and}
  \bibinfo{author}{\bibfnamefont{P.}~\bibnamefont{Navratil}},
  \bibinfo{journal}{Phys. Rev. C} \textbf{\bibinfo{volume}{94}},
  \bibinfo{pages}{011301(R)} (\bibinfo{year}{2016}),
  \urlprefix\url{https://journals.aps.org/prc/abstract/10.1103/PhysRevC.94.011301}.

\bibitem[{\citenamefont{Hagen et~al.}(2016)\citenamefont{Hagen, Jansen, and
  Papenbrock}}]{Hagen2016}
\bibinfo{author}{\bibfnamefont{G.}~\bibnamefont{Hagen}},
  \bibinfo{author}{\bibfnamefont{G.~R.} \bibnamefont{Jansen}},
  \bibnamefont{and}
  \bibinfo{author}{\bibfnamefont{T.}~\bibnamefont{Papenbrock}},
  \bibinfo{journal}{Phys. Rev. Lett.} \textbf{\bibinfo{volume}{117}},
  \bibinfo{pages}{172501} (\bibinfo{year}{2016}), ISSN
  \bibinfo{issn}{10797114}.

\bibitem[{\citenamefont{Stroberg et~al.}(2017)\citenamefont{Stroberg, Calci,
  Hergert, Holt, Bogner, Roth, and Schwenk}}]{Stroberg2017}
\bibinfo{author}{\bibfnamefont{S.~R.} \bibnamefont{Stroberg}},
  \bibinfo{author}{\bibfnamefont{A.}~\bibnamefont{Calci}},
  \bibinfo{author}{\bibfnamefont{H.}~\bibnamefont{Hergert}},
  \bibinfo{author}{\bibfnamefont{J.~D.} \bibnamefont{Holt}},
  \bibinfo{author}{\bibfnamefont{S.~K.} \bibnamefont{Bogner}},
  \bibinfo{author}{\bibfnamefont{R.}~\bibnamefont{Roth}}, \bibnamefont{and}
  \bibinfo{author}{\bibfnamefont{A.}~\bibnamefont{Schwenk}},
  \bibinfo{journal}{Phys. Rev. Lett.} \textbf{\bibinfo{volume}{118}},
  \bibinfo{pages}{032502} (\bibinfo{year}{2017}),
  \urlprefix\url{https://journals.aps.org/prl/abstract/10.1103/PhysRevLett.118.032502}.

\bibitem[{\citenamefont{Morris et~al.}(2018)\citenamefont{Morris, Simonis,
  Stroberg, Stumpf, Hagen, Holt, Jansen, Papenbrock, Roth, and
  Schwenk}}]{Morris2018}
\bibinfo{author}{\bibfnamefont{T.~D.} \bibnamefont{Morris}},
  \bibinfo{author}{\bibfnamefont{J.}~\bibnamefont{Simonis}},
  \bibinfo{author}{\bibfnamefont{S.~R.} \bibnamefont{Stroberg}},
  \bibinfo{author}{\bibfnamefont{C.}~\bibnamefont{Stumpf}},
  \bibinfo{author}{\bibfnamefont{G.}~\bibnamefont{Hagen}},
  \bibinfo{author}{\bibfnamefont{J.~D.} \bibnamefont{Holt}},
  \bibinfo{author}{\bibfnamefont{G.~R.} \bibnamefont{Jansen}},
  \bibinfo{author}{\bibfnamefont{T.}~\bibnamefont{Papenbrock}},
  \bibinfo{author}{\bibfnamefont{R.}~\bibnamefont{Roth}}, \bibnamefont{and}
  \bibinfo{author}{\bibfnamefont{A.}~\bibnamefont{Schwenk}},
  \bibinfo{journal}{Phys. Rev. Lett.} \textbf{\bibinfo{volume}{120}},
  \bibinfo{pages}{152503} (\bibinfo{year}{2018}), ISSN
  \bibinfo{issn}{10797114}.

\bibitem[{\citenamefont{Gysbers et~al.}(2019)\citenamefont{Gysbers, Hagen,
  Holt, Jansen, Morris, Navr{\'{a}}til, Papenbrock, Quaglioni, Schwenk,
  Stroberg et~al.}}]{Gysbers2019}
\bibinfo{author}{\bibfnamefont{P.}~\bibnamefont{Gysbers}},
  \bibinfo{author}{\bibfnamefont{G.}~\bibnamefont{Hagen}},
  \bibinfo{author}{\bibfnamefont{J.~D.} \bibnamefont{Holt}},
  \bibinfo{author}{\bibfnamefont{G.~R.} \bibnamefont{Jansen}},
  \bibinfo{author}{\bibfnamefont{T.~D.} \bibnamefont{Morris}},
  \bibinfo{author}{\bibfnamefont{P.}~\bibnamefont{Navr{\'{a}}til}},
  \bibinfo{author}{\bibfnamefont{T.}~\bibnamefont{Papenbrock}},
  \bibinfo{author}{\bibfnamefont{S.}~\bibnamefont{Quaglioni}},
  \bibinfo{author}{\bibfnamefont{A.}~\bibnamefont{Schwenk}},
  \bibinfo{author}{\bibfnamefont{S.~R.} \bibnamefont{Stroberg}},
  \bibnamefont{et~al.}, \bibinfo{journal}{Nature Physics}
  \textbf{\bibinfo{volume}{15}}, \bibinfo{pages}{428} (\bibinfo{year}{2019}),
  ISSN \bibinfo{issn}{17452481}.

\bibitem[{\citenamefont{Garnsworthy et~al.}(2019)\citenamefont{Garnsworthy,
  Svensson, Bowry, Dunlop, MacLean, Olaizola, Smith, Ali, Andreoiu, Ash
  et~al.}}]{Garnsworthy2019}
\bibinfo{author}{\bibfnamefont{A.~B.} \bibnamefont{Garnsworthy}},
  \bibinfo{author}{\bibfnamefont{C.~E.} \bibnamefont{Svensson}},
  \bibinfo{author}{\bibfnamefont{M.}~\bibnamefont{Bowry}},
  \bibinfo{author}{\bibfnamefont{R.}~\bibnamefont{Dunlop}},
  \bibinfo{author}{\bibfnamefont{A.~D.} \bibnamefont{MacLean}},
  \bibinfo{author}{\bibfnamefont{B.}~\bibnamefont{Olaizola}},
  \bibinfo{author}{\bibfnamefont{J.~K.} \bibnamefont{Smith}},
  \bibinfo{author}{\bibfnamefont{F.~A.} \bibnamefont{Ali}},
  \bibinfo{author}{\bibfnamefont{C.}~\bibnamefont{Andreoiu}},
  \bibinfo{author}{\bibfnamefont{J.~E.} \bibnamefont{Ash}},
  \bibnamefont{et~al.}, \bibinfo{journal}{Nucl. Instrum. Methods Phys. Res.
  Sect. A} \textbf{\bibinfo{volume}{918}}, \bibinfo{pages}{9 }
  (\bibinfo{year}{2019}), ISSN \bibinfo{issn}{0168-9002},
  \urlprefix\url{http://www.sciencedirect.com/science/article/pii/S0168900218317662}.

\bibitem[{\citenamefont{Burrows}(2007)}]{Burrows2007}
\bibinfo{author}{\bibfnamefont{T.~W.} \bibnamefont{Burrows}},
  \bibinfo{journal}{Nucl. Data Sheets} \textbf{\bibinfo{volume}{108}},
  \bibinfo{pages}{923} (\bibinfo{year}{2007}), ISSN \bibinfo{issn}{00903752},
  \urlprefix\url{http://www.sciencedirect.com/science/article/pii/S0090375207000403}.

\bibitem[{\citenamefont{Warburton et~al.}(1970)\citenamefont{Warburton,
  Alburger, and Engelbertink}}]{Warburton1970}
\bibinfo{author}{\bibfnamefont{E.~K.} \bibnamefont{Warburton}},
  \bibinfo{author}{\bibfnamefont{D.~E.} \bibnamefont{Alburger}},
  \bibnamefont{and} \bibinfo{author}{\bibfnamefont{G.~A.~P.}
  \bibnamefont{Engelbertink}}, \bibinfo{journal}{Phys. Rev. C}
  \textbf{\bibinfo{volume}{2}}, \bibinfo{pages}{1427} (\bibinfo{year}{1970}),
  ISSN \bibinfo{issn}{0556-2813},
  \urlprefix\url{http://link.aps.org/doi/10.1103/PhysRevC.2.1427}.

\bibitem[{\citenamefont{Alburger et~al.}(1984)\citenamefont{Alburger,
  Warburton, and Brown}}]{Alburger1984}
\bibinfo{author}{\bibfnamefont{D.~E.} \bibnamefont{Alburger}},
  \bibinfo{author}{\bibfnamefont{E.~K.} \bibnamefont{Warburton}},
  \bibnamefont{and} \bibinfo{author}{\bibfnamefont{B.~A.} \bibnamefont{Brown}},
  \bibinfo{journal}{Phys. Rev. C} \textbf{\bibinfo{volume}{30}},
  \bibinfo{pages}{1005} (\bibinfo{year}{1984}), ISSN \bibinfo{issn}{0556-2813},
  \urlprefix\url{http://link.aps.org/doi/10.1103/PhysRevC.30.1005}.

\bibitem[{\citenamefont{Hanspal et~al.}(1985)\citenamefont{Hanspal, Clarke,
  Griffiths, Karban, and Roman}}]{Hanspal1985}
\bibinfo{author}{\bibfnamefont{J.~S.} \bibnamefont{Hanspal}},
  \bibinfo{author}{\bibfnamefont{N.~M.} \bibnamefont{Clarke}},
  \bibinfo{author}{\bibfnamefont{R.~J.} \bibnamefont{Griffiths}},
  \bibinfo{author}{\bibfnamefont{O.}~\bibnamefont{Karban}}, \bibnamefont{and}
  \bibinfo{author}{\bibfnamefont{S.}~\bibnamefont{Roman}},
  \bibinfo{journal}{Nucl. Phys. A} \textbf{\bibinfo{volume}{436}},
  \bibinfo{pages}{236} (\bibinfo{year}{1985}), ISSN \bibinfo{issn}{03759474},
  \urlprefix\url{http://www.sciencedirect.com/science/article/pii/0375947485901988}.

\bibitem[{\citenamefont{Martin et~al.}(1972)\citenamefont{Martin, Buenerd,
  Dupont, and Chabre}}]{Martin1972}
\bibinfo{author}{\bibfnamefont{P.}~\bibnamefont{Martin}},
  \bibinfo{author}{\bibfnamefont{M.}~\bibnamefont{Buenerd}},
  \bibinfo{author}{\bibfnamefont{Y.}~\bibnamefont{Dupont}}, \bibnamefont{and}
  \bibinfo{author}{\bibfnamefont{M.}~\bibnamefont{Chabre}},
  \bibinfo{journal}{Nucl. Phys. A} \textbf{\bibinfo{volume}{185}},
  \bibinfo{pages}{465} (\bibinfo{year}{1972}), ISSN \bibinfo{issn}{03759474},
  \urlprefix\url{http://www.sciencedirect.com/science/article/pii/0375947472900255}.

\bibitem[{\citenamefont{Williams-Norton and Abegg}(1977)}]{Williams-Norton1977}
\bibinfo{author}{\bibfnamefont{M.~E.} \bibnamefont{Williams-Norton}}
  \bibnamefont{and} \bibinfo{author}{\bibfnamefont{R.}~\bibnamefont{Abegg}},
  \bibinfo{journal}{Nucl. Phys. A} \textbf{\bibinfo{volume}{291}},
  \bibinfo{pages}{429} (\bibinfo{year}{1977}), ISSN \bibinfo{issn}{03759474},
  \urlprefix\url{http://www.sciencedirect.com/science/article/pii/037594747790330X}.

\bibitem[{\citenamefont{Kuroyanagi et~al.}(1964)\citenamefont{Kuroyanagi,
  Tamura, Tanaka, and Morinaga}}]{Kuroyanagi1964}
\bibinfo{author}{\bibfnamefont{T.}~\bibnamefont{Kuroyanagi}},
  \bibinfo{author}{\bibfnamefont{T.}~\bibnamefont{Tamura}},
  \bibinfo{author}{\bibfnamefont{K.}~\bibnamefont{Tanaka}}, \bibnamefont{and}
  \bibinfo{author}{\bibfnamefont{H.}~\bibnamefont{Morinaga}},
  \bibinfo{journal}{Nucl. Phys.} \textbf{\bibinfo{volume}{50}},
  \bibinfo{pages}{417} (\bibinfo{year}{1964}), ISSN \bibinfo{issn}{00295582},
  \urlprefix\url{http://www.sciencedirect.com/science/article/pii/0029558264902184}.

\bibitem[{\citenamefont{Fortier et~al.}(1978)\citenamefont{Fortier, Hourani,
  Rao, and Gal{\`{e}}s}}]{Fortier1978}
\bibinfo{author}{\bibfnamefont{S.}~\bibnamefont{Fortier}},
  \bibinfo{author}{\bibfnamefont{E.}~\bibnamefont{Hourani}},
  \bibinfo{author}{\bibfnamefont{M.~N.} \bibnamefont{Rao}}, \bibnamefont{and}
  \bibinfo{author}{\bibfnamefont{S.}~\bibnamefont{Gal{\`{e}}s}},
  \bibinfo{journal}{Nucl. Phys. A} \textbf{\bibinfo{volume}{311}},
  \bibinfo{pages}{324} (\bibinfo{year}{1978}), ISSN \bibinfo{issn}{03759474},
  \urlprefix\url{http://linkinghub.elsevier.com/retrieve/pii/0375947478905183}.

\bibitem[{\citenamefont{Kovar et~al.}(1978)\citenamefont{Kovar, Henning,
  Zeidman, Eisen, Erskine, Fortune, Ophel, Sperr, and Vigdor}}]{Kovar1978}
\bibinfo{author}{\bibfnamefont{D.~G.} \bibnamefont{Kovar}},
  \bibinfo{author}{\bibfnamefont{W.}~\bibnamefont{Henning}},
  \bibinfo{author}{\bibfnamefont{B.}~\bibnamefont{Zeidman}},
  \bibinfo{author}{\bibfnamefont{Y.}~\bibnamefont{Eisen}},
  \bibinfo{author}{\bibfnamefont{J.~R.} \bibnamefont{Erskine}},
  \bibinfo{author}{\bibfnamefont{H.~T.} \bibnamefont{Fortune}},
  \bibinfo{author}{\bibfnamefont{T.~R.} \bibnamefont{Ophel}},
  \bibinfo{author}{\bibfnamefont{P.}~\bibnamefont{Sperr}}, \bibnamefont{and}
  \bibinfo{author}{\bibfnamefont{S.~E.} \bibnamefont{Vigdor}},
  \bibinfo{journal}{Phys. Rev. C} \textbf{\bibinfo{volume}{17}},
  \bibinfo{pages}{83} (\bibinfo{year}{1978}), ISSN \bibinfo{issn}{0556-2813},
  \urlprefix\url{http://journals.aps.org/prc/abstract/10.1103/PhysRevC.17.83}.

\bibitem[{\citenamefont{Humanic et~al.}(1982)\citenamefont{Humanic, Ernst,
  Henning, and Zeidman}}]{Humanic1982}
\bibinfo{author}{\bibfnamefont{T.~J.} \bibnamefont{Humanic}},
  \bibinfo{author}{\bibfnamefont{H.}~\bibnamefont{Ernst}},
  \bibinfo{author}{\bibfnamefont{W.}~\bibnamefont{Henning}}, \bibnamefont{and}
  \bibinfo{author}{\bibfnamefont{B.}~\bibnamefont{Zeidman}},
  \bibinfo{journal}{Phys. Rev. C} \textbf{\bibinfo{volume}{26}},
  \bibinfo{pages}{993} (\bibinfo{year}{1982}), ISSN \bibinfo{issn}{0556-2813},
  \urlprefix\url{http://journals.aps.org/prc/abstract/10.1103/PhysRevC.26.993}.

\bibitem[{\citenamefont{Petersen et~al.}(1983)\citenamefont{Petersen, Ascuitto,
  Kovar, Henning, and Sanders}}]{Petersen1983}
\bibinfo{author}{\bibfnamefont{J.~F.} \bibnamefont{Petersen}},
  \bibinfo{author}{\bibfnamefont{R.~J.} \bibnamefont{Ascuitto}},
  \bibinfo{author}{\bibfnamefont{D.~G.} \bibnamefont{Kovar}},
  \bibinfo{author}{\bibfnamefont{W.}~\bibnamefont{Henning}}, \bibnamefont{and}
  \bibinfo{author}{\bibfnamefont{S.~J.} \bibnamefont{Sanders}},
  \bibinfo{journal}{Phys. Rev. C} \textbf{\bibinfo{volume}{28}},
  \bibinfo{pages}{710} (\bibinfo{year}{1983}), ISSN \bibinfo{issn}{0556-2813},
  \urlprefix\url{http://journals.aps.org/prc/abstract/10.1103/PhysRevC.28.710}.

\bibitem[{\citenamefont{Mando et~al.}(1983)\citenamefont{Mando, Sona, Taccetti,
  Brandolini, Alvarez, and Poli}}]{Mando1983}
\bibinfo{author}{\bibfnamefont{P.~A.} \bibnamefont{Mando}},
  \bibinfo{author}{\bibfnamefont{P.}~\bibnamefont{Sona}},
  \bibinfo{author}{\bibfnamefont{N.}~\bibnamefont{Taccetti}},
  \bibinfo{author}{\bibfnamefont{F.}~\bibnamefont{Brandolini}},
  \bibinfo{author}{\bibfnamefont{C.~R.} \bibnamefont{Alvarez}},
  \bibnamefont{and} \bibinfo{author}{\bibfnamefont{M.~D.} \bibnamefont{Poli}},
  \bibinfo{journal}{J. Phys. G Nucl. Phys.} \textbf{\bibinfo{volume}{9}},
  \bibinfo{pages}{1211} (\bibinfo{year}{1983}), ISSN \bibinfo{issn}{0305-4616},
  \urlprefix\url{http://iopscience.iop.org/article/10.1088/0305-4616/9/10/009}.

\bibitem[{\citenamefont{Strauch et~al.}(2000)\citenamefont{Strauch, {von
  Neumann-Cosel}, Rangacharyulu, Richter, Schrieder, Schweda, and
  Wambach}}]{Strauch2000}
\bibinfo{author}{\bibfnamefont{S.}~\bibnamefont{Strauch}},
  \bibinfo{author}{\bibfnamefont{P.}~\bibnamefont{{von Neumann-Cosel}}},
  \bibinfo{author}{\bibfnamefont{C.}~\bibnamefont{Rangacharyulu}},
  \bibinfo{author}{\bibfnamefont{A.}~\bibnamefont{Richter}},
  \bibinfo{author}{\bibfnamefont{G.}~\bibnamefont{Schrieder}},
  \bibinfo{author}{\bibfnamefont{K.}~\bibnamefont{Schweda}}, \bibnamefont{and}
  \bibinfo{author}{\bibfnamefont{J.}~\bibnamefont{Wambach}},
  \bibinfo{journal}{Phys. Rev. Lett.} \textbf{\bibinfo{volume}{85}},
  \bibinfo{pages}{2913} (\bibinfo{year}{2000}), ISSN \bibinfo{issn}{0031-9007},
  \urlprefix\url{http://journals.aps.org/prl/abstract/10.1103/PhysRevLett.85.2913}.

\bibitem[{\citenamefont{Broda}(2001)}]{Broda2001}
\bibinfo{author}{\bibfnamefont{R.}~\bibnamefont{Broda}}, \bibinfo{journal}{Acta
  Phys. Pol. B} \textbf{\bibinfo{volume}{32}}, \bibinfo{pages}{2577}
  (\bibinfo{year}{2001}),
  \urlprefix\url{http://www.actaphys.uj.edu.pl/fulltext?series=Reg&vol=32&page=2577}.

\bibitem[{\citenamefont{Schweda et~al.}(2001)\citenamefont{Schweda, Carter,
  Cowley, Diesener, Fearick, F{\"{o}}rtsch, Lawrie, von Neumann-Cosel, Pilcher,
  Richter et~al.}}]{Schweda2001}
\bibinfo{author}{\bibfnamefont{K.}~\bibnamefont{Schweda}},
  \bibinfo{author}{\bibfnamefont{J.}~\bibnamefont{Carter}},
  \bibinfo{author}{\bibfnamefont{A.~A.} \bibnamefont{Cowley}},
  \bibinfo{author}{\bibfnamefont{H.}~\bibnamefont{Diesener}},
  \bibinfo{author}{\bibfnamefont{R.~W.} \bibnamefont{Fearick}},
  \bibinfo{author}{\bibfnamefont{S.~V.} \bibnamefont{F{\"{o}}rtsch}},
  \bibinfo{author}{\bibfnamefont{J.~J.} \bibnamefont{Lawrie}},
  \bibinfo{author}{\bibfnamefont{P.}~\bibnamefont{von Neumann-Cosel}},
  \bibinfo{author}{\bibfnamefont{J.~V.} \bibnamefont{Pilcher}},
  \bibinfo{author}{\bibfnamefont{A.}~\bibnamefont{Richter}},
  \bibnamefont{et~al.}, \bibinfo{journal}{Phys. Lett. B}
  \textbf{\bibinfo{volume}{506}}, \bibinfo{pages}{247} (\bibinfo{year}{2001}),
  ISSN \bibinfo{issn}{03702693},
  \urlprefix\url{http://www.sciencedirect.com/science/article/pii/S0370269301002301}.

\bibitem[{\citenamefont{Montanari et~al.}(2012)\citenamefont{Montanari, Leoni,
  Mengoni, Valiente-Dobon, Benzoni, Blasi, Bocchi, Bortignon, Bottoni, Bracco
  et~al.}}]{Montanari2012}
\bibinfo{author}{\bibfnamefont{D.}~\bibnamefont{Montanari}},
  \bibinfo{author}{\bibfnamefont{S.}~\bibnamefont{Leoni}},
  \bibinfo{author}{\bibfnamefont{D.}~\bibnamefont{Mengoni}},
  \bibinfo{author}{\bibfnamefont{J.~J.} \bibnamefont{Valiente-Dobon}},
  \bibinfo{author}{\bibfnamefont{G.}~\bibnamefont{Benzoni}},
  \bibinfo{author}{\bibfnamefont{N.}~\bibnamefont{Blasi}},
  \bibinfo{author}{\bibfnamefont{G.}~\bibnamefont{Bocchi}},
  \bibinfo{author}{\bibfnamefont{P.~F.} \bibnamefont{Bortignon}},
  \bibinfo{author}{\bibfnamefont{S.}~\bibnamefont{Bottoni}},
  \bibinfo{author}{\bibfnamefont{A.}~\bibnamefont{Bracco}},
  \bibnamefont{et~al.}, \bibinfo{journal}{Phys. Rev. C}
  \textbf{\bibinfo{volume}{85}}, \bibinfo{pages}{044301}
  (\bibinfo{year}{2012}), ISSN \bibinfo{issn}{0556-2813},
  \urlprefix\url{http://link.aps.org/doi/10.1103/PhysRevC.85.044301}.

\bibitem[{\citenamefont{Crawford et~al.}(2017)\citenamefont{Crawford,
  Macchiavelli, Fallon, Albers, Bader, Bazin, Campbell, Clark, Cromaz, Dilling
  et~al.}}]{Crawford2017}
\bibinfo{author}{\bibfnamefont{H.~L.} \bibnamefont{Crawford}},
  \bibinfo{author}{\bibfnamefont{A.~O.} \bibnamefont{Macchiavelli}},
  \bibinfo{author}{\bibfnamefont{P.}~\bibnamefont{Fallon}},
  \bibinfo{author}{\bibfnamefont{M.}~\bibnamefont{Albers}},
  \bibinfo{author}{\bibfnamefont{V.~M.} \bibnamefont{Bader}},
  \bibinfo{author}{\bibfnamefont{D.}~\bibnamefont{Bazin}},
  \bibinfo{author}{\bibfnamefont{C.~M.} \bibnamefont{Campbell}},
  \bibinfo{author}{\bibfnamefont{R.~M.} \bibnamefont{Clark}},
  \bibinfo{author}{\bibfnamefont{M.}~\bibnamefont{Cromaz}},
  \bibinfo{author}{\bibfnamefont{J.}~\bibnamefont{Dilling}},
  \bibnamefont{et~al.}, \bibinfo{journal}{Phys. Rev. C}
  \textbf{\bibinfo{volume}{95}}, \bibinfo{pages}{064317:}
  (\bibinfo{year}{2017}), ISSN \bibinfo{issn}{2469-9985},
  \urlprefix\url{http://link.aps.org/doi/10.1103/PhysRevC.95.064317}.

\bibitem[{\citenamefont{Bjerregaard et~al.}(1965)\citenamefont{Bjerregaard,
  Hansen, and Sidenius}}]{Bjerregaard1965}
\bibinfo{author}{\bibfnamefont{J.~H.} \bibnamefont{Bjerregaard}},
  \bibinfo{author}{\bibfnamefont{O.}~\bibnamefont{Hansen}}, \bibnamefont{and}
  \bibinfo{author}{\bibfnamefont{G.}~\bibnamefont{Sidenius}},
  \bibinfo{journal}{Phys. Rev.} \textbf{\bibinfo{volume}{138}},
  \bibinfo{pages}{B1097} (\bibinfo{year}{1965}), ISSN
  \bibinfo{issn}{0031-899X},
  \urlprefix\url{http://link.aps.org/doi/10.1103/PhysRev.138.B1097}.

\bibitem[{\citenamefont{Belote et~al.}(1966)\citenamefont{Belote, Chen, Hansen,
  and Rapaport}}]{Belote1966}
\bibinfo{author}{\bibfnamefont{T.~A.} \bibnamefont{Belote}},
  \bibinfo{author}{\bibfnamefont{H.~Y.} \bibnamefont{Chen}},
  \bibinfo{author}{\bibfnamefont{O.}~\bibnamefont{Hansen}}, \bibnamefont{and}
  \bibinfo{author}{\bibfnamefont{J.}~\bibnamefont{Rapaport}},
  \bibinfo{journal}{Phys. Rev.} \textbf{\bibinfo{volume}{142}},
  \bibinfo{pages}{624} (\bibinfo{year}{1966}), ISSN \bibinfo{issn}{0031-899X},
  \urlprefix\url{http://link.aps.org/doi/10.1103/PhysRev.142.624}.

\bibitem[{\citenamefont{Cranston et~al.}(1970)\citenamefont{Cranston, Birkett,
  White, and Hughes}}]{Cranston1970}
\bibinfo{author}{\bibfnamefont{F.~P.} \bibnamefont{Cranston}},
  \bibinfo{author}{\bibfnamefont{R.~E.} \bibnamefont{Birkett}},
  \bibinfo{author}{\bibfnamefont{D.~H.} \bibnamefont{White}}, \bibnamefont{and}
  \bibinfo{author}{\bibfnamefont{J.~A.} \bibnamefont{Hughes}},
  \bibinfo{journal}{Nucl. Phys. A} \textbf{\bibinfo{volume}{153}},
  \bibinfo{pages}{413} (\bibinfo{year}{1970}), ISSN \bibinfo{issn}{03759474},
  \urlprefix\url{http://www.sciencedirect.com/science/article/pii/0375947470907803}.

\bibitem[{\citenamefont{Wang et~al.}(2017)\citenamefont{Wang, Audi, Kondev,
  Huang, Naimi, and Xu}}]{Wang2017}
\bibinfo{author}{\bibfnamefont{M.}~\bibnamefont{Wang}},
  \bibinfo{author}{\bibfnamefont{G.}~\bibnamefont{Audi}},
  \bibinfo{author}{\bibfnamefont{F.~G.} \bibnamefont{Kondev}},
  \bibinfo{author}{\bibfnamefont{W.~J.} \bibnamefont{Huang}},
  \bibinfo{author}{\bibfnamefont{S.}~\bibnamefont{Naimi}}, \bibnamefont{and}
  \bibinfo{author}{\bibfnamefont{X.}~\bibnamefont{Xu}},
  \bibinfo{journal}{Chinese Physics C} \textbf{\bibinfo{volume}{41}}
  (\bibinfo{year}{2017}), ISSN \bibinfo{issn}{16741137}.

\bibitem[{\citenamefont{Bylinskii and Craddock}(2013)}]{Bylinskii2013}
\bibinfo{author}{\bibfnamefont{I.}~\bibnamefont{Bylinskii}} \bibnamefont{and}
  \bibinfo{author}{\bibfnamefont{M.~K.} \bibnamefont{Craddock}},
  \bibinfo{journal}{Hyperfine Interact.} \textbf{\bibinfo{volume}{225}},
  \bibinfo{pages}{9} (\bibinfo{year}{2013}), ISSN \bibinfo{issn}{0304-3843},
  \urlprefix\url{http://link.springer.com/10.1007/s10751-013-0878-6}.

\bibitem[{\citenamefont{Svensson and Garnsworthy}(2013)}]{Svensson2013}
\bibinfo{author}{\bibfnamefont{C.~E.} \bibnamefont{Svensson}} \bibnamefont{and}
  \bibinfo{author}{\bibfnamefont{A.~B.} \bibnamefont{Garnsworthy}},
  \bibinfo{journal}{Hyperfine Interact.} \textbf{\bibinfo{volume}{225}},
  \bibinfo{pages}{127} (\bibinfo{year}{2013}), ISSN \bibinfo{issn}{0304-3843},
  \urlprefix\url{http://link.springer.com/10.1007/s10751-013-0889-3}.

\bibitem[{\citenamefont{Rizwan et~al.}(2016)\citenamefont{Rizwan, Garnsworthy,
  Andreoiu, Ball, Chester, Domingo, Dunlop, Hackman, Rand, Smith
  et~al.}}]{Rizwan2016}
\bibinfo{author}{\bibfnamefont{U.}~\bibnamefont{Rizwan}},
  \bibinfo{author}{\bibfnamefont{A.~B.} \bibnamefont{Garnsworthy}},
  \bibinfo{author}{\bibfnamefont{C.}~\bibnamefont{Andreoiu}},
  \bibinfo{author}{\bibfnamefont{G.~C.} \bibnamefont{Ball}},
  \bibinfo{author}{\bibfnamefont{A.}~\bibnamefont{Chester}},
  \bibinfo{author}{\bibfnamefont{T.}~\bibnamefont{Domingo}},
  \bibinfo{author}{\bibfnamefont{R.}~\bibnamefont{Dunlop}},
  \bibinfo{author}{\bibfnamefont{G.}~\bibnamefont{Hackman}},
  \bibinfo{author}{\bibfnamefont{E.~T.} \bibnamefont{Rand}},
  \bibinfo{author}{\bibfnamefont{J.~K.} \bibnamefont{Smith}},
  \bibnamefont{et~al.}, \bibinfo{journal}{Nucl. Instrum. Methods Phys. Res.
  Sect. A} \textbf{\bibinfo{volume}{820}}, \bibinfo{pages}{126}
  (\bibinfo{year}{2016}), ISSN \bibinfo{issn}{01689002},
  \urlprefix\url{http://www.sciencedirect.com/science/article/pii/S0168900216300341}.

\bibitem[{\citenamefont{Garnsworthy et~al.}(2017)\citenamefont{Garnsworthy,
  Pearson, Bishop, Shaw, Smith, Bowry, Bildstein, Hackman, Garrett, Linn
  et~al.}}]{Garnsworthy2017}
\bibinfo{author}{\bibfnamefont{A.~B.} \bibnamefont{Garnsworthy}},
  \bibinfo{author}{\bibfnamefont{C.~J.} \bibnamefont{Pearson}},
  \bibinfo{author}{\bibfnamefont{D.}~\bibnamefont{Bishop}},
  \bibinfo{author}{\bibfnamefont{B.}~\bibnamefont{Shaw}},
  \bibinfo{author}{\bibfnamefont{J.~K.} \bibnamefont{Smith}},
  \bibinfo{author}{\bibfnamefont{M.}~\bibnamefont{Bowry}},
  \bibinfo{author}{\bibfnamefont{V.}~\bibnamefont{Bildstein}},
  \bibinfo{author}{\bibfnamefont{G.}~\bibnamefont{Hackman}},
  \bibinfo{author}{\bibfnamefont{P.~E.} \bibnamefont{Garrett}},
  \bibinfo{author}{\bibfnamefont{Y.}~\bibnamefont{Linn}}, \bibnamefont{et~al.},
  \bibinfo{journal}{Nucl. Instrum. Methods Phys. Res. Sect. A}
  \textbf{\bibinfo{volume}{853}}, \bibinfo{pages}{85} (\bibinfo{year}{2017}),
  \urlprefix\url{https://www.sciencedirect.com/science/article/pii/S0168900217302243}.

\bibitem[{\citenamefont{Dunlop et~al.}(2015)}]{TPeak}
\bibinfo{author}{\bibfnamefont{R.}~\bibnamefont{Dunlop}} \bibnamefont{et~al.}
  (\bibinfo{year}{2015}),
  \urlprefix\url{https://github.com/GRIFFINCollaboration/GRSISort/wiki/TPeak}.

\bibitem[{\citenamefont{Grinyer et~al.}(2005)\citenamefont{Grinyer, Svensson,
  Andreoiu, Andreyev, Austin, Ball, Chakrawarthy, Finlay, Garrett, Hackman
  et~al.}}]{Grinyer2005}
\bibinfo{author}{\bibfnamefont{G.~F.} \bibnamefont{Grinyer}},
  \bibinfo{author}{\bibfnamefont{C.~E.} \bibnamefont{Svensson}},
  \bibinfo{author}{\bibfnamefont{C.}~\bibnamefont{Andreoiu}},
  \bibinfo{author}{\bibfnamefont{A.~N.} \bibnamefont{Andreyev}},
  \bibinfo{author}{\bibfnamefont{R.~A.~E.} \bibnamefont{Austin}},
  \bibinfo{author}{\bibfnamefont{G.~C.} \bibnamefont{Ball}},
  \bibinfo{author}{\bibfnamefont{R.~S.} \bibnamefont{Chakrawarthy}},
  \bibinfo{author}{\bibfnamefont{P.}~\bibnamefont{Finlay}},
  \bibinfo{author}{\bibfnamefont{P.~E.} \bibnamefont{Garrett}},
  \bibinfo{author}{\bibfnamefont{G.}~\bibnamefont{Hackman}},
  \bibnamefont{et~al.}, \bibinfo{journal}{Phys. Rev. C}
  \textbf{\bibinfo{volume}{71}}, \bibinfo{pages}{044309}
  (\bibinfo{year}{2005}), ISSN \bibinfo{issn}{1089490X},
  \urlprefix\url{https://journals.aps.org/prc/abstract/10.1103/PhysRevC.71.044309}.

\bibitem[{\citenamefont{Hardy et~al.}(1977)\citenamefont{Hardy, Carraz, Jonson,
  and Hansen}}]{Hardy1977}
\bibinfo{author}{\bibfnamefont{J.~C.} \bibnamefont{Hardy}},
  \bibinfo{author}{\bibfnamefont{L.~C.} \bibnamefont{Carraz}},
  \bibinfo{author}{\bibfnamefont{B.}~\bibnamefont{Jonson}}, \bibnamefont{and}
  \bibinfo{author}{\bibfnamefont{P.~G.} \bibnamefont{Hansen}},
  \bibinfo{journal}{Phys. Lett. B} \textbf{\bibinfo{volume}{71}},
  \bibinfo{pages}{307} (\bibinfo{year}{1977}), ISSN \bibinfo{issn}{03702693},
  \urlprefix\url{http://www.sciencedirect.com/science/article/pii/0370269377902234}.

\bibitem[{\citenamefont{Smith et~al.}(2019)\citenamefont{Smith, MacLean,
  Ashfield, Chester, Garnsworthy, and Svensson}}]{Smith2018}
\bibinfo{author}{\bibfnamefont{J.~K.} \bibnamefont{Smith}},
  \bibinfo{author}{\bibfnamefont{A.~D.} \bibnamefont{MacLean}},
  \bibinfo{author}{\bibfnamefont{W.}~\bibnamefont{Ashfield}},
  \bibinfo{author}{\bibfnamefont{A.}~\bibnamefont{Chester}},
  \bibinfo{author}{\bibfnamefont{A.~B.} \bibnamefont{Garnsworthy}},
  \bibnamefont{and} \bibinfo{author}{\bibfnamefont{C.~E.}
  \bibnamefont{Svensson}}, \bibinfo{journal}{Nucl. Instrum. Meth. Phys. Res.
  Sect. A} \textbf{\bibinfo{volume}{922}}, \bibinfo{pages}{47 }
  (\bibinfo{year}{2019}), ISSN \bibinfo{issn}{0168-9002},
  \urlprefix\url{http://www.sciencedirect.com/science/article/pii/S0168900218314116}.

\bibitem[{\citenamefont{Endt}(1979)}]{Endt1979}
\bibinfo{author}{\bibfnamefont{P.~M.} \bibnamefont{Endt}},
  \bibinfo{journal}{At. Data Nucl. Data Tables} \textbf{\bibinfo{volume}{23}},
  \bibinfo{pages}{547} (\bibinfo{year}{1979}),
  \urlprefix\url{http://www.sciencedirect.com/science/article/pii/0092640X79900305{\%}5Cnpapers2://publication/uuid/331D72DF-6400-443E-BED7-9EE1FE145EA0}.

\bibitem[{\citenamefont{Endt}(1993)}]{Endt1993}
\bibinfo{author}{\bibfnamefont{P.~M.} \bibnamefont{Endt}},
  \bibinfo{journal}{At. Data Nucl. Data Tables} \textbf{\bibinfo{volume}{55}},
  \bibinfo{pages}{171} (\bibinfo{year}{1993}), ISSN \bibinfo{issn}{0092640X}.

\bibitem[{\citenamefont{Multhauf et~al.}(1975)\citenamefont{Multhauf, Tirsell,
  Raman, and McGrory}}]{Multhauf1975}
\bibinfo{author}{\bibfnamefont{L.~G.} \bibnamefont{Multhauf}},
  \bibinfo{author}{\bibfnamefont{K.~G.} \bibnamefont{Tirsell}},
  \bibinfo{author}{\bibfnamefont{S.}~\bibnamefont{Raman}}, \bibnamefont{and}
  \bibinfo{author}{\bibfnamefont{J.~B.} \bibnamefont{McGrory}},
  \bibinfo{journal}{Physics Letters B} \textbf{\bibinfo{volume}{57}},
  \bibinfo{pages}{44} (\bibinfo{year}{1975}), ISSN \bibinfo{issn}{03702693}.

\bibitem[{\citenamefont{Huck et~al.}(1981)\citenamefont{Huck, Klotz, Knipper,
  Mieh{\'{e}}, Richard-Serre, and Walter}}]{Huck1981}
\bibinfo{author}{\bibfnamefont{A.}~\bibnamefont{Huck}},
  \bibinfo{author}{\bibfnamefont{G.}~\bibnamefont{Klotz}},
  \bibinfo{author}{\bibfnamefont{A.}~\bibnamefont{Knipper}},
  \bibinfo{author}{\bibfnamefont{C.}~\bibnamefont{Mieh{\'{e}}}},
  \bibinfo{author}{\bibfnamefont{C.}~\bibnamefont{Richard-Serre}},
  \bibnamefont{and} \bibinfo{author}{\bibfnamefont{G.}~\bibnamefont{Walter}},
  \bibinfo{journal}{Proc. Int. Conf. Nuclei Far from Stability}
  \textbf{\bibinfo{volume}{2}}, \bibinfo{pages}{378} (\bibinfo{year}{1981}),
  \urlprefix\url{http://cds.cern.ch/record/134702?ln=en}.

\bibitem[{\citenamefont{Carraz et~al.}(1982)\citenamefont{Carraz, Hansen, Huck,
  Jonson, Klotz, Knipper, Kratz, Mi{\'{e}}h{\'{e}}, Mattsson, Nyman
  et~al.}}]{Carraz1982}
\bibinfo{author}{\bibfnamefont{L.~C.} \bibnamefont{Carraz}},
  \bibinfo{author}{\bibfnamefont{P.~G.} \bibnamefont{Hansen}},
  \bibinfo{author}{\bibfnamefont{A.}~\bibnamefont{Huck}},
  \bibinfo{author}{\bibfnamefont{B.}~\bibnamefont{Jonson}},
  \bibinfo{author}{\bibfnamefont{G.}~\bibnamefont{Klotz}},
  \bibinfo{author}{\bibfnamefont{A.}~\bibnamefont{Knipper}},
  \bibinfo{author}{\bibfnamefont{K.~L.} \bibnamefont{Kratz}},
  \bibinfo{author}{\bibfnamefont{C.}~\bibnamefont{Mi{\'{e}}h{\'{e}}}},
  \bibinfo{author}{\bibfnamefont{S.}~\bibnamefont{Mattsson}},
  \bibinfo{author}{\bibfnamefont{G.}~\bibnamefont{Nyman}},
  \bibnamefont{et~al.}, \bibinfo{journal}{Physics Letters B}
  \textbf{\bibinfo{volume}{109}}, \bibinfo{pages}{419} (\bibinfo{year}{1982}),
  ISSN \bibinfo{issn}{03702693}.

\end{thebibliography}

\newpage

\end{document}